\documentclass[%
 reprint,
superscriptaddress,
twocolumn,
 amsmath,amssymb,
 aps,
prb,
]{revtex4-2}

\usepackage{booktabs}
\usepackage{graphicx}
\usepackage{dcolumn}
\usepackage{bm}
\usepackage{xcolor}
\usepackage{float}
\usepackage{microtype}
\usepackage[normalem]{ulem}

\usepackage{hyperref}
\hypersetup{colorlinks=true,breaklinks,linkcolor=blue,urlcolor=blue,citecolor=blue}
\usepackage{orcidlink}

\newcommand\AddAuthorComment[3]{
    {\color{#1} ({\bf #2}%
        \if\relax\detokenize{#3}\relax%
        \else%
            {\normalfont: #3}%
        \fi%
    )}%
}
\newcommand\AuthorReplace[5]{
    \AddAuthorComment{#1}{#2}{#3}%
    \if\relax\detokenize{#4#5}\relax%
    \else%
        { \color{#1}\sout{#4}\uwave{#5}}
    \fi%
}

\definecolor{darkgreen}{rgb}{0,0.5,0}

\newcommand\Dmax{\ensuremath{D_\mathrm{max}}}
\newcommand{\cL}{\ensuremath{\mathcal{L}}}
\newcommand{\order}{\ensuremath{\mathcal{O}}}
\newcommand{\ph}{\ensuremath{\textit{ph}}}
\newcommand{\pp}{\ensuremath{\textit{pp}}}
\newcommand\phbar{\ensuremath{\overline{\textit{ph}}}}
\renewcommand\vec{\boldsymbol}
\newcommand{\ms}{\ensuremath{\phantom{-}}}

\begin{document}

\preprint{APS/123-QED}

\title{Two-particle calculations with quantics tensor trains: Solving the parquet equations}

\author{Stefan Rohshap\,\orcidlink{0009-0007-2953-8831}}
\email{stefan.rohshap@tuwien.ac.at}
\affiliation{%
    Institute of Solid State Physics, TU Wien, 1040 Vienna, Austria}%

\author{Marc K.\ Ritter\,\orcidlink{0000-0002-2960-5471}}
\email{ritter.marc@lmu.de}
\affiliation{%
    Arnold Sommerfeld Center for Theoretical Physics, Center for NanoScience, and Munich Center for Quantum Science and Technology, Ludwig-Maximilians-Universität München, 80333 Munich, Germany}%

\author{Hiroshi Shinaoka\,\orcidlink{0000-0002-7058-8765}}
\affiliation{%
    Department of Physics, Saitama University, Saitama 338-8570, Japan}%

\author{Jan von Delft\,\orcidlink{0000-0002-8655-0999}}
\affiliation{%
    Arnold Sommerfeld Center for Theoretical Physics, Center for NanoScience, and Munich Center for Quantum Science and Technology, Ludwig-Maximilians-Universität München, 80333 Munich, Germany}%

\author{Markus Wallerberger\,\orcidlink{0000-0002-9992-1541}}
\affiliation{%
    Institute of Solid State Physics, TU Wien, 1040 Vienna, Austria}%

\author{Anna Kauch\,\orcidlink{0000-0002-7669-0090}}
\email{kauch@ifp.tuwien.ac.at}
\affiliation{%
    Institute of Solid State Physics, TU Wien, 1040 Vienna, Austria}%

\date{\today}

\begin{abstract}
We present the first application of quantics tensor trains (QTTs) and tensor cross interpolation (TCI) to the solution of a full set of self-consistent equations for multivariate functions, the so-called parquet equations. We show that the steps needed to evaluate the equations (Bethe--Salpeter equations, parquet equation and Schwinger--Dyson equation) can be decomposed into basic operations on the QTT-TCI (QTCI) compressed objects. The repeated application of these operations does not lead to a loss of accuracy beyond a specified tolerance and the iterative scheme converges even for numerically demanding parameters. 
As examples we take the Hubbard model in the atomic limit and the single impurity Anderson model, where the basic objects in parquet equations, the two-particle vertices, depend on three frequencies, but not on momenta.
The results show that this approach is able to overcome major computational bottlenecks of standard numerical methods. 
The applied methods allow for an exponential increase of the number of grid points included in the calculations, and a corresponding exponential reduction of the computational error, for a linear increase in computational cost.
\end{abstract}

\maketitle

\section{\label{sec:introduction}Introduction}


The understanding of many important excitations of electronic systems -- magnons, excitons, or other composite objects -- requires understanding correlations at the two particle level.
Two-particle quantities -- correlation functions or scattering amplitudes (vertices) --  are inherently large objects, with multiple dependencies: If we consider scattering of two particles, the amplitude will depend on the energies, momenta and spin-orbitals of two incoming and two outgoing particles. The number of independent variables can be reduced using conservation laws, but each independent spin-orbital combination still depends on three momenta and three frequencies. Numerical representation of these multivariate functions on uniform grids is very expensive due to the third power scaling of memory in the number of discrete momenta or energies. On the other hand, large ranges are required to faithfully represent complicated structures which the vertices show in all their dependencies \cite{Rohringer2012,Lee2021,Krien2022}. When the vertices are themselves variables in diagrammatic equations, as it is the case in parquet equations~\cite{Dominicis64,Dominicis64-2, Bickers04}, the required computation time becomes prohibitive \cite{Tam2013}. Several solutions to this problem have been proposed so far,  either based on partial reduction of the number of frequency and/or momentum variables that need to be treated on grids \cite{Eckhardt2020, Astretsov2019, Wentzell2020,Krien2020a, Krien2020b} or based on compact representation of the frequency dependence  in a suitable basis \cite{Wallerberger2021,Shinaoka2020,Shinaoka2018, Kiese2024}. The former still do not lead to true dimensional reduction of the full parquet equations problem. The latter are very promising and provide another path to dimensional reduction, alternative to the one described in this work. Recently, also a wavelet-based decomposition for efficiently compressing two-particle quantities has been proposed~\cite{Mallat1989,Daubechies1990,Moghadas2024}.

In this paper, we present the first full computation of the self-consistent solution of parquet equations in the quantics tensor train representation. This representation, based on length/energy scale separation, leads to significant dimensional reduction of the problem, removing memory bottlenecks. The computational cost becomes logarithmic in grid size and depends strongly only  on the maximum bond dimension, which is small enough in many physics applications. Hence, the overall computational cost is significantly reduced.

The quantics tensor train (QTT) representation of multivariate functions has already been around for a decade or so \cite{Oseledets2009,Khoromskij2011,Dolgov2012,Qttbook}, but it was only recently applied to various fields of natural science such as turbulence \cite{Gourianov2022-vn,peddinti2023complete, kornev2023,holscher2024, Gourianov2024}, plasma physics \cite{Ye2022}, quantum chemistry \cite{Jolly2023}, and quantum field theory of the many electron problem \cite{Shinaoka2023}.
For quantum field theories, the QTT representation provides a compact representation of the space-time dependence of the correlation functions~\cite{Shinaoka2023}.
First many-body calculations of Feynman diagrams with the QTT representation in imaginary time \cite{Ishida2024} and in nonequilibrium \cite{Murray2024} already show the potential of the method. A very favorable scaling of the QTT representation with temperature has been conjectured in Ref.~\onlinecite{takahashi2024compactnessquanticstensortrain}.
In parallel, the tensor cross interpolation (TCI) method was applied to evaluations of diagrams in many-body physics \cite{NunezFernandez2022, Erpenbeck2023,NunezFernandez2024,Eckstein2024}.
TCI can be combined with the quantics tensor train representation to form QTT+TCI=QTCI, a powerful approach with diverse applications \cite{Ritter2024}.

In order to apply QTCI to parquet equations, we break down these equations (Bethe--Salpeter equation, parquet equation and Schwinger--Dyson equation) into basic operations on QTTs, represented by matrix product operators (MPOs). We use MPO-MPO contractions for matrix and elementwise multiplications and construct a new MPO for affine transformations that are needed to perform channel transformations (variable shifts) occurring in the parquet equation.
Our approach scales as \(\order(\Dmax^4 R)\), with maximum bond dimension of \(\Dmax\) and grid size \(2^{3R}\). The computational cost is only logarithmic in grid size and the main bottleneck is shifted to the maximum bond dimension. We have verified this scaling in two benchmark models: the Hubbard atom and the single impurity Anderson model (SIAM). In both models, the two-particle vertices are fully dynamical (dependent on three frequencies) but local (independent of momentum).
In both cases, we empirically find that an overall accuracy $< 10^{-3}$ of the full self-consistent solution
can be achieved with a bond dimension up to 200 
even for challenging parameters close to a divergence line in the Hubbard atom.



This paper is organized as follows. In Sec.~\ref{sec:models} we introduce the concrete Hamiltonians (Hubbard atom and single impurity Anderson model) for which we will present the results. Further, in Sec. \ref{sec:parquet-equations} we first recall definitions of one- and two-particle Green's functions and vertices and set the notation used in the paper. Then we provide in detail the full set of  parquet equations that we solve. Additional information on the equations and notation is also provided in App.~\ref{app:fermionic-frequencies}. In \ref{sec:quantics-tensor-trains} we introduce quantics tensor trains, the tensor cross interpolation method and matrix product operators.
These techniques are used to construct efficient implementations of the parquet equations in Sec.~\ref{sec:parquet-in-qtt} and Apps.~\ref{app:mpo-affine}-\ref{app:channel-transformation}.
We also provide results for the compression of the vertices and scaling of the bond dimension for each of the operations needed to complete one loop of parquet equations in Sec.~\ref{sec:parquet-in-qtt}. More details and additional plots can be found in Appendices \ref{app:qtt-parquet}-\ref{app:qtt-sde}. Next, in Sec.~\ref{sec:results}, we show results for the full self-consistent iterative scheme and its technical limitations (with details also in App.~\ref{app:technical-limitations}). In the last section~\ref{sec:conclusions}, we conclude and provide outlook.

\section{Models}
\label{sec:models}

In the current work we  focus on the solution of equations for two-particle vertices in the (Matsubara) frequency space. Although in general the vertices are also dependent on momentum and orbital degrees of freedom, we limit ourselves to simple models for which the vertices depend only on frequency but not on momentum.
We present results for two benchmark models:
the Hubbard atom, where exact analytical expressions for the vertex functions are known \cite{Thunstroem2018}
and the single impurity Anderson model~\cite{Hewson93}, where high-quality numerical data is available ~\cite{Krien2022}.
The treatment of the frequency dependence of vertices presented in this work can be directly extended to models with additional orbital and momentum dependencies. The possibility of such extensions will be discussed in App. \ref{app:extensions}.

\subsection{Hubbard atom}

The Hubbard atom is an extreme simplification of the Hubbard model in which the hopping amplitudes of the electrons between sites are put to zero. Although this is a drastic change, the Hubbard atom represents many of the features of the strong-coupling limit of the Hubbard model \cite{Thunstroem2018}.  Without hopping, each atom is independent and described by the following Hamiltonian:
\begin{align} \label{eq:hamiltonian-hubbard-atom}
    \hat{\mathcal{H}} =U \hat{n}_{\uparrow} \hat{n}_{\downarrow} - \mu(\hat{n}_{ \uparrow}+ \hat{n}_{\downarrow}),
\end{align}
with $\hat n_{\sigma} = \hat{c}_{\sigma}^{\dagger} \hat{c}_{\sigma}$ and the fermionic annihilation (creation) operator $\hat{c}_{\sigma}^{(\dagger)}$ that annihilates (creates) an electron with spin $\sigma$. The on-site Coulomb repulsion between two electrons is given by $U$ and the chemical potential is set to $\mu = \frac{U}{2}$ (half-filling). The only other energy scale beside $U$ in this model is the temperature $T$, which we define in the same units as $U$, setting $k_B\equiv 1$ and $\hbar\equiv 1$.

\subsection{Single-impurity Anderson model}
In the single-impurity Anderson model (SIAM), the interacting atom is not isolated, but coupled to a bath of non-interacting electrons. The SIAM Hamiltonian is~\cite{Hewson93}
\begin{align}
    \hat{\mathcal{H}} &= \sum_{\vec k \sigma} \varepsilon_{\vec{k}}\hat{c}_{\vec{k}, \sigma}^{\dagger} \hat{c}_{\vec{k},\sigma}^{\phantom\dagger} + \sum_{\vec{k} \sigma} \left( V_{\vec{k}} \hat{c}_{\vec{k}, \sigma}^{\dagger} \hat{d}_{\sigma}^{\phantom\dagger} + V^{*}_{\vec{k}} \hat{d}_{\sigma}^{\dagger} \hat{c}_{\vec{k}, \sigma}^{\phantom\dagger} \right)\nonumber \\
    &+ U \hat{n}_{d, \uparrow} \hat{n}_{d, \downarrow} + \varepsilon_d (\hat{n}_{d,  \uparrow}+ \hat{n}_{d, \downarrow}), 
\end{align}
where the impurity is described by the fermionic annihilation (creation) operators $\hat{d}_{\sigma}^{(\dagger)}$, the number operator $\hat{n}_{d,\sigma} = \hat{d}_{\sigma}^{\dagger} \hat{d}_{\sigma}$, the impurity one-particle energy level $\varepsilon_d$ and the onsite repulsion $U$. The bath is described by the kinetic term $\hat{c}_{\vec{k}, \sigma}^{(\dagger)}$ with one-particle energies $\varepsilon_{\vec{k}}$. The hybridization between the impurity and the bath  is given by $V_{\vec{k}}$. The bath parameters jointly determine the frequency dependent hybridization function:
\begin{align}
  \Delta(\nu)  = \sum_{\vec{k}} \frac{|V_{\vec{k}}|^2}{i\nu - \varepsilon_{\vec{k}}}.
\end{align}
In this work we use the following hybridization function 
\begin{align}
    \Delta (\nu) = -\frac{i V^2}{D} \arctan\left( \frac{D}{\nu} \right),
    \label{eq:Deltaflat}
\end{align}
which corresponds to a flat density of states of the bath electrons $\rho (\epsilon) = \theta(D-|\epsilon|)/(2D)$ with bandwidth $D$ and $V_{\vec{k}} = V$. We will present results for $V=2$, $D=10$ and half-filling, i.e. with $\varepsilon_d=-U/2$. 

\section{Parquet equations}
\label{sec:parquet-equations}
The parquet equations are a set of exact  relations between different classes of two-particle vertices and between the self-energy and the full two-particle vertex~\cite{Dominicis64, Dominicis64-2}. A good introduction to the formalism is provided in~\cite{Bickers04}. Here we only recall the equations, using the notation of Refs.~\cite{Rohringer2012, Rohringer18,Held2022,victory2019}. Before we introduce the equations themselves, we first recapitulate some definitions in order to set the notation. 

\subsection{One-particle quantities}
The one-particle Green's function in the Matsubara frequency space $G_{\sigma}({\nu})$ is defined as the Fourier transform of the (imaginary-time ordered) two-point correlation function:
\begin{align}
    G_{\sigma}(\nu) = -\int_0^{\beta}\!\!d\tau e^{i \nu \tau} \langle T_{\tau} \hat{c}_{ \sigma}(\tau) \hat{c}_{\sigma}^{\dagger}(0) \rangle \,,
\end{align}
with $\tau$ denoting the imaginary time, $\beta\equiv 1/T$ the inverse temperature, and $\nu = (2n+1)\pi / \beta, n \in \mathbb{Z}$ denoting the (discrete) fermionic Matsubara frequencies. Through the Dyson equation, we further define the self-energy $\Sigma_{\sigma}(\nu)$ 
\begin{align}
   G_{\sigma}(\nu) = \frac{1}{G^{-1}_{0,\sigma}(\nu)- \Sigma_{\sigma}(\nu)}
   \,, \label{eq:dyson}
\end{align}
where $G_{0,\sigma}(\nu)$ is the Green's function of the noninteracting system:
\begin{align}
    G_{0,\sigma}(\nu) = \frac{1}{i\nu + \frac{U}{2} - \Delta (\nu)} \,,
\end{align}
where we have set the following model-dependent parameters to values corresponding to half filling:
For the Hubbard atom, \(\Delta(\nu)\) vanishes, and the chemical potential is \(\mu = U/2\). For the single-impurity Anderson model, we set the chemical potential to \(\mu = 0\) and the one-particle energy level to \(\epsilon_d = -U/2\).

\subsection{Two-particle quantities}

The two-particle Green's function in Matsubara frequencies is the Fourier transform of the  (imaginary time ordered) four-point correlator
\begin{align}
    G_{\sigma_1\ldots\sigma_4}^{\nu_1\nu_2\nu_3} = &\int_0^{\beta}\!\!\!d\tau_1 \int_0^{\beta}\!\!\!d\tau_2\int_0^{\beta}\!\!\!d\tau_3\;e^{i \nu_1 \tau_1+i\nu_2 \tau_2+i \nu_3 \tau_3 
    }\nonumber \\ \times & \langle T_{\tau} \hat{c}_{ \sigma_1}(\tau_1) \hat{c}_{\sigma_2}^{\dagger}(\tau_2)\hat{c}_{ \sigma_3}(\tau_3) \hat{c}_{\sigma_4}^{\dagger}(0) \rangle.
    \label{eq:G2}
\end{align}
In the above definition of the Fourier transform, the two-particle Green's function is dependent on three fermionic Matsubara frequencies $\nu_1,\nu_2, \nu_3$. 
In the context of Bethe--Salpeter equations (defined below in Sec.~\ref{sec:bse_eq}), it is more convenient to parameterize two-particle quantities as a function of two fermionic frequencies $\nu, \nu'$ and one bosonic Matsubara frequency $\omega=\frac{2n\pi}{\beta}, n \in \mathbb{Z}$.
There are three important conventions of this parametrization: the particle-hole (\ph) channel notation, where $\omega=\nu_1+\nu_2$; the particle-particle (\pp) channel notation, where $\omega=\nu_1+\nu_3$, and the transversal particle-hole (\phbar) channel notation, where $\omega=\nu_2+\nu_3$.
In the parquet approach, it is necessary to transform between these conventions using so-called channel transformations outlined in App.~\ref{app:fermionic-frequencies}. The reason, as we will see in Sec.~\ref{sec:parquet}, is that the  parquet equation  mixes vertex functions that are represented in different frequency channel parametrizations.


In this work we use the $\mathrm{SU}(2)$ symmetry of the discussed models, which allows us, together with spin conservation, to reduce the number of spin components that need to be computed to the following:
$G_{\sigma\sigma\sigma'\sigma'}$, which we will denote by $G_{\sigma\sigma'}$, and $G_{\sigma(-\sigma)(-\sigma)\sigma}$, which can be shown to be equal to $G_{\sigma\sigma} - G_{\sigma(-\sigma)}$. Furthermore, since $G_{\sigma\sigma'} = G_{(-\sigma)(-\sigma')}$,
we only need to compute $G_{\uparrow\uparrow}$ and $G_{\uparrow\downarrow}$.
We will proceed in a similar manner for the vertex $F$ (see below).
From here on we will also drop the spin index from the one-particle objects $G_0$, $G$ and $\Sigma$, since $G_\uparrow=G_\downarrow$.

The full two-particle vertex $F$ is the connected part of the two-particle Green's function with ``amputated legs''. In the particle-hole channel it is related to the two-particle Green's function through
\begin{align}
    G_{\sigma\sigma'}^{\nu\nu'\omega} &=
    G(\nu) G(\nu')\delta_{\omega 0} - G(\nu) G(\nu+\omega)\delta_{\nu \nu'}\delta_{\sigma\sigma'} 
    \nonumber \\
    & \quad -
    G(\nu) G({\nu+\omega}){F}_{\sigma\sigma'}^{\nu \nu'\omega}G(\nu')G(\nu'+\omega).
    \label{eq:Fdef}
\end{align}
Apart from channels stemming from different frequency parametrizations (\ph, \pp, and \phbar), it is convenient to introduce also linear combinations of spin components. The following spin combinations will be used for vertices in the \ph\ frequency channel:
\begin{align}
    F_d &= F_{\uparrow \uparrow} +F_{\uparrow \downarrow}, \nonumber \\
    F_m &= F_{\uparrow \uparrow} -F_{\uparrow \downarrow}, \label{eq:spins}
\end{align}
which physically correspond to the density (d) and magnetic (m) spin components. 

The same vertex $F$ can be represented in the particle-particle channel frequency parametrization: $F^{\pp}$ (see Appendix \ref{app:fermionic-frequencies} for details). In the \pp\ channel, the convenient spin combinations are the following:
\begin{align}
    F_s &= F^{\pp}_{\uparrow \uparrow} -F^{\pp}_{\uparrow \downarrow}, \nonumber \\
    F_t &= F^{\pp}_{\uparrow \uparrow},
\end{align}
physically corresponding to the singlet (s) and triplet (t) spin components. In the following, we will predominantly use the spin component notation, i.e. $d/m/s/t$, assuming that the $d$ or $m$ spin components are always in the \ph\ frequency channel notation and the $s$ or $t$ spin components are always in the \pp\ frequency channel notation.

As we will see later (in Sec.~\ref{sec:bse_eq}), in the above four spin combinations the Bethe-Salpeter equations decouple in the spin variable.


\subsection{Bethe--Salpeter equations}
\label{sec:bse_eq}
The full vertex $F$ contains all diagrams irrespective of their two-particle reducibility. The Bethe--Salpeter equations relate the full two-particle vertex to sets of two-particle irreducible diagrams. This is analogous to Dyson's equation \eqref{eq:dyson}, however, 
in the two-particle case the notion of irreducibility is not unique. Instead of one Dyson equation, we have independent Bethe--Salpeter equations (BSEs) in particle-hole and particle-particle channels. In the \ph\ channel, we have the equations for density and magnetic components:
\begin{subequations}%
\begin{align}
    &F_{d}^{\nu \nu' \omega} = \Gamma_{d}^{\nu \nu' \omega} - \frac{1}{\beta^2} \sum_{\nu_1 \nu_2} \Gamma_{d}^{\nu \nu_1 \omega} \chi_{0,\ph}^{\nu_1 \nu_2 \omega} F_{d}^{\nu_2 \nu' \omega}, \label{eq:bse-density} \\
    & F_{m}^{\nu \nu' \omega} = \Gamma_{m}^{\nu \nu' \omega} - \frac{1}{\beta^2} \sum_{\nu_1 \nu_2} \Gamma_{m}^{\nu \nu_1 \omega} \chi_{0,\ph}^{\nu_1 \nu_2 \omega} F_{m}^{\nu_2 \nu' \omega};
\end{align}
in the \pp\ channel, we have the equations for the singlet and triplet components: 
\begin{align}
      & F_{s}^{\nu \nu' \omega} = \Gamma_{s}^{\nu \nu' \omega} + \frac{1}{\beta^2} \sum_{\nu_1 \nu_2} F_{s}^{\nu \nu_1 \omega} \chi_{0,\pp}^{\nu_1 \nu_2 \omega} \Gamma_{s}^{\nu_2 \nu' \omega},\\
    & F_{t}^{\nu \nu' \omega} = \Gamma_{t}^{\nu \nu' \omega} - \frac{1}{\beta^2} \sum_{\nu_1 \nu_2} F_{t}^{\nu \nu_1 \omega} \chi_{0,\pp}^{\nu_1 \nu_2 \omega} \Gamma_{t}^{\nu_2 \nu' \omega}. 
  \label{eq:bse-triplet} 
\end{align}
\label{eq:bse}%
\end{subequations}
The above equations define four irreducible vertices $\Gamma_r$, $r=d/m/s/t$ that are irreducible in either \ph\ channel ($r=d/m$) or \pp\ channel ($r=s/t$).
We define vertices reducible in these channels simply as:
\begin{subequations}
\begin{align}
    \Phi_{d/m}^{\nu \nu' \omega} &=  F_{d/m}^{\nu \nu' \omega}  - \Gamma_{d/m}^{\nu \nu' \omega},
    \\
    \Phi_{s/t}^{\nu \nu' \omega} &= F_{s/t}^{\nu \nu' \omega} - \Gamma_{s/t}^{\nu \nu' \omega}. 
\end{align}
\label{eq:phi}%
\end{subequations}
The $\chi_0$'s are products of two one-particle Green's functions and are defined as follows:
\begin{subequations}
\begin{align}
    &\chi_{0,\ph}^{\nu \nu' \omega}=-\beta G(\nu)G(\nu+\omega) \delta_{\nu \nu'},\\
    &\chi_{0,\pp}^{\nu \nu' \omega}=-\frac{\beta}{2}G(\nu)G(-\nu-\omega) \delta_{\nu \nu'}.
\end{align}
\end{subequations}
The pair propagators (also called  bare generalized susceptibilities) $\chi_0$'s are diagonal in $\nu,\nu'$, which means that the sum in Eqs. \eqref{eq:bse} runs over only one fermionic Matsubara frequency index. For convenience of actual numerical evaluations, we however keep the double fermionic frequency dependence in $\chi_0$'s.

{Due to convenient parametrization of the frequency dependence of the vertices, i.e. the \ph\ channel for $d/m$ and \pp\ channel for $s/t$, the BSEs~\eqref{eq:bse} are diagonal both in the bosonic frequency $\omega$ and in the spin components $d/m/s/t$.} 

\subsection{\label{sec:parquet}Parquet equation}

Through Eqs.~\eqref{eq:bse}-\eqref{eq:phi} we  defined reducible vertices $\Phi_{d/m}$ and $\Phi_{s/t}$ in \ph\ and \pp\  channels, respectively. These vertices correspond to different physical processes that happen in the \ph\ and \pp\ scattering channels and are generated by the BSEs~\eqref{eq:bse}. The parquet equation mixes these processes, allowing for balance between contributions generated by all of the  BSEs~\eqref{eq:bse}. 
In a simplified way, the parquet equation can be represented as the following sum of terms:\begin{align}
F = \Lambda + \Phi_{\ph} + \Phi_{\phbar} + \Phi_{\pp},
\label{eq:parquet}
\end{align}
where $\Phi_{\ph}$ denotes contributions coming from $\Phi_d$ or $\Phi_m$, $\Phi_{\phbar}$ contributions coming from $\Phi_{d/m}$, but in the \phbar\ frequency parametrization, and $\Phi_{\pp}$ contributions from $\Phi_s$ or $\Phi_t$ (more details can be found in App.~\ref{app:fermionic-frequencies} or in~Ref.~\onlinecite{Bickers04}).
The first summand, $\Lambda$, contains so-called fully two-particle irreducible diagrams, i.e., contributions which cannot be generated by the two-particle BSEs. 

Since in the BSEs the reducible vertices $\Phi_r$ are in different frequency channel parametrizations, in order to sum the contributions we have to transform them into a common parametrization. The explicit form of the parquet equation for $F_d$ is then the following~\cite{Rohringer18}:
\begin{align}
    F_d^{\nu \nu' \omega} &= \Lambda_d^{\nu \nu' \omega} +  \Phi_d^{\nu \nu' \omega} -\tfrac{1}{2} \Phi_d^{\nu (\nu+\omega) (\nu'-\nu)}
    \nonumber\\
    &\quad- \tfrac{3}{2} \Phi_m^{\nu (\nu+\omega) (\nu'-\nu)}
    + \tfrac{1}{2} \Phi_s^{\nu \nu' (-\omega -\nu -\nu')}
    \nonumber\\
    &\quad+ \tfrac{3}{2} \Phi_t^{\nu \nu' (-\omega -\nu -\nu')}
    \,,
    \label{eq:parquet-density} 
\end{align}
where $\Lambda_d$ is the density component of the fully irreducible vertex. \(\Lambda_d\) cannot be obtained from the BSEs and has to be provided from outside the parquet scheme. In the examples presented in Sec.~\ref{sec:results}, we either use the exact expression (it is known for the Hubbard atom) or we use a weak coupling approximation for it.
Eq.~\eqref{eq:parquet} can be used to generate equations analogous to Eq.~\eqref{eq:parquet-density} for $F_m$, $F_s$, and $F_t$. We provide them explicitly in App. \ref{app:fermionic-frequencies}.

\subsection{Schwinger--Dyson equation}

The last equation that belongs to the set of parquet equations is the Schwinger--Dyson equation (SDE) that relates the two-particle vertex $F$ to the self-energy
\begin{align}
    \Sigma (\nu) = \frac{Un}{2} - \frac{U}{\beta^2} \sum_{\nu' \omega} F_{\uparrow\downarrow }^{\nu \nu' \omega} G(\nu') G(\nu'+\omega) G(\nu+\omega)
    \,,\label{eq:sde}
\end{align}
where $n$ is the average particle density and \(F_{ \uparrow\downarrow}^{\nu \nu' \omega} = \frac{1}{2}(F_{d}^{\nu \nu' \omega} - F_{m}^{\nu \nu' \omega})\) (follows from Eq. \eqref{eq:spins}).

\subsection{Iterative parquet scheme}
\label{sec:iterparquet}

Assuming we know the fully irreducible vertex $\Lambda$ (or have a good approximation for it) the parquet equation \eqref{eq:parquet} together with the BSEs \eqref{eq:bse} and the SDE \eqref{eq:sde}, as well as the Dyson equation \eqref{eq:dyson}, applied iteratively, will generate all the vertices for the model given by the Hamiltonian and $G_0$: \{$F_{d/m}$, $F_{s/t}$, $\Gamma_{d/m}$, $\Gamma_{s/t}$, $\Phi_{d/m}$, $\Phi_{s/t}$\} as well as the self-energy $\Sigma$ and the Green's function $G$. The iterations are repeated until the difference between consecutive values falls below a given tolerance.


The input quantity that does not change in the iterations, namely the fully irreducible vertex $\Lambda$, is known exactly (even analytically \cite{Thunstroem2018}) for the Hubbard atom and we use this exact expression. For the SIAM we use the so-called parquet approximation which sets $\Lambda$ equal to the bare interaction $U$ (it is the lowest order diagram in $\Lambda$, the next order appearing in the diagrammatic expansion of $\Lambda$ is $U^4$). Explicitly written out in spin components, the parquet approximation reads:
\begin{align}
\Lambda_d =U, \quad \Lambda_m = -U, \quad  \Lambda_s = 2U, \quad  \Lambda_t = 0.    
\end{align}

\begin{figure}[t]
    \centering 
    \includegraphics[width=0.28\textwidth]{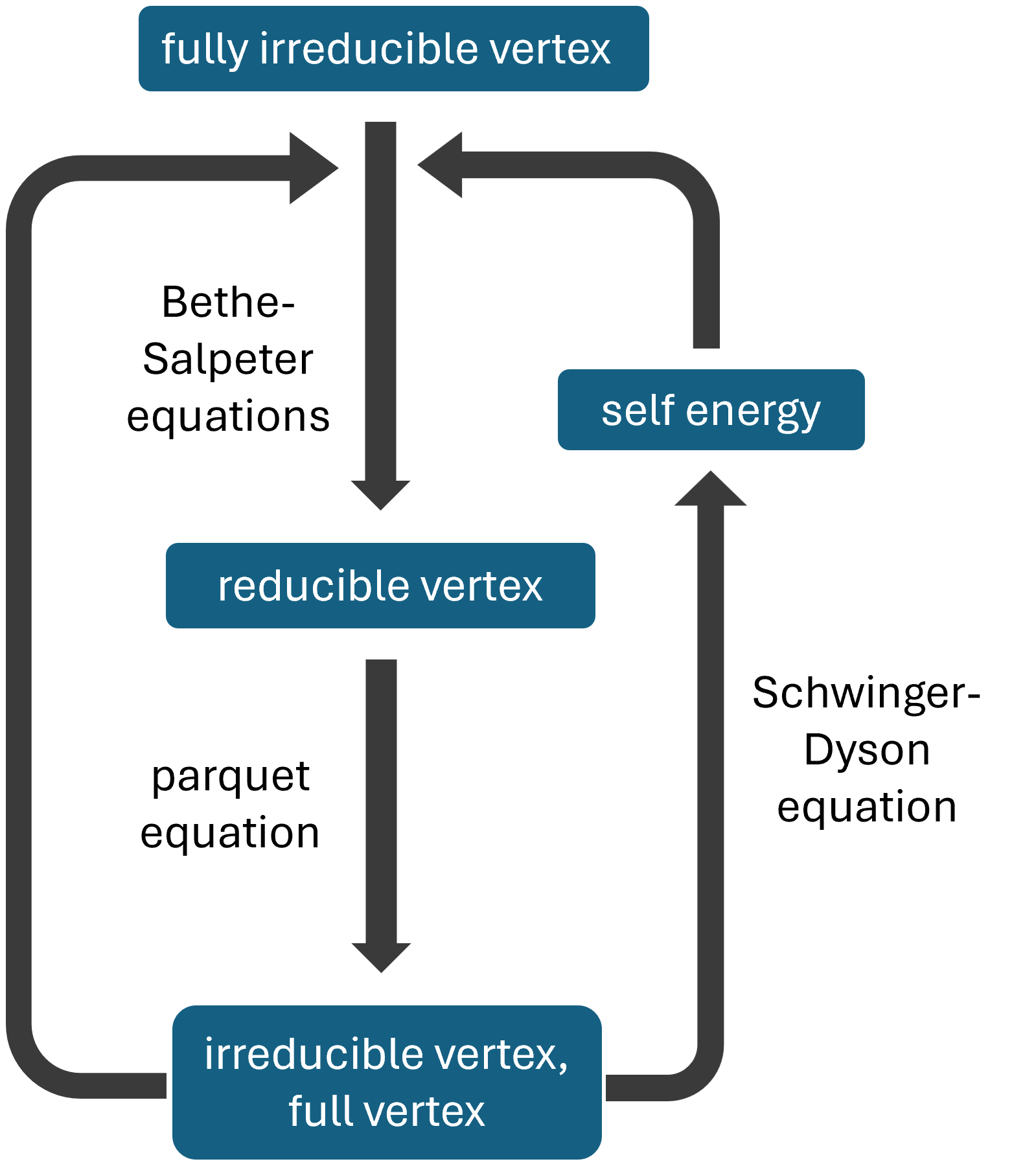}
    \caption{Iterative parquet scheme.}
    \label{fig:iterative-parquet}
\end{figure}

In Fig. \ref{fig:iterative-parquet} we show how the actual iterated loop is implemented in this work. In the first cycle of the iteration, we either set $\Gamma_r = \Lambda_r, F_d = U,  F_m = -U, F_s = 2U,  F_t = \Lambda_t$ or use previous results for smaller values of $\beta U$. 
After initialization, the BSEs~\eqref{eq:bse} are evaluated and the reducible vertices $\Phi_r$ are obtained. Then, from the parquet equation~\eqref{eq:parquet-density}  (for all channels) the new $F$ is computed and the irreducible vertices $\Gamma_r$ are updated from~\eqref{eq:phi}. From the new $F$ the self-energy $\Sigma$ can be obtained from SDE \eqref{eq:sde} and the one-particle Green's function from \eqref{eq:dyson}. The self-energy update does not have to happen in each iteration -- depending on the value of $U$, it might be faster to update it every 5 or 10 iterations. With updated $F$, $\Gamma_r$, and $G$ the cycle consisting of the nine equations is repeated until convergence is reached. In this work, in order to keep things simple, we limit convergence acceleration to a linear mixing update to the reducible vertex, \(\Phi_{n+1} = \alpha \Phi'_n + (1-\alpha) \Phi_{n}\) with mixing parameter \(\alpha\), where \(\Phi_n\) is the reducible vertex in iteration \(n\) and \(\Phi'_n\) is the result of applying a single cycle to \(\Phi_n\). 

We conclude this section with some comments on convergence issues. 
First, the iterative parquet scheme is not guaranteed to converge. Second,  there are cases where the iterative scheme leads to a false solution. This is the case, e.g., for the Hubbard atom for large $\beta U$ beyond the first divergence of the irreducible vertex~\cite{Kozik2015}. In this work we do not address such cases and focus on examples and parameter regimes where the parquet scheme should converge to the physical solution. Nevertheless, convergence can also be influenced by other factors. The first numerical solution for the Hubbard model on a $4\times4$ cluster~\cite{Tam2013} showed the difficulty of achieving convergence, particularly when the crossing symmetry (see App.~\ref{app:fermionic-frequencies}) was not obeyed. An important factor in improving stability of the iterative scheme turned out to be the inclusion of vertex asymptotics, i.e., a prediction for values of the vertex that fall beyond the frequency range used, either by introducing so-called kernels~\cite{Li16,victory2019, Wentzell2020,Eckhardt2020}, or by removing the asymptotic parts of vertices altogether in the single-boson exchange reformulation~\cite{Krien2020b, Krien2022}. In this work we do not use asymptotics and do not suffer from convergence problems mainly because we are able to use very large grids. In the future it might be important to include asymptotic, as in Refs.~\cite{Li16,victory2019, Wentzell2020,Eckhardt2020} or~\cite{Krien2020b, Krien2022}, to improve stability but also to avoid some technical issues that are addressed in App.~\ref{app:technical-limitations}.

\section{Quantics Tensor trains}
\label{sec:quantics-tensor-trains}

Numerical solution of the iterative parquet scheme introduced in the previous section suffers from the curse of dimensionality: the vertex functions have multiple frequency (and in general also momentum) arguments. Discretizing these multivariate functions on naive grids requires a number of grid points that grows exponentially with the number of function arguments, which therefore becomes very expensive already for a moderate number of grid points in each argument.  The solution to this problem that we propose is (i) to represent each variable through a set of binary numbers (hence `quantics') corresponding to different length/energy scales; (ii) to factorize the dependence on each argument at each length/energy scale into a tensor train (TT), also known as matrix product state (MPS). If the problem has some kind of scale separation, the resulting quantics tensor train (QTT) is expected to have a small maximum bond dimension. Since this is the case in many physical problems, such an approach is potentially very powerful. It has already been shown to reduce computational costs significantly in several applications with high-dimensional functions \cite{Shinaoka2023,Ritter2024, Oseledets2009, Oseledets2011, Khoromskij2011}.

{In this section, we introduce the definition of the QTT representation, then present a method for efficient compression of multivariate functions into a QTT, namely the tensor cross interpolation (TCI). Finally,  we also introduce matrix product operators (MPOs) that are needed for computations with the QTTs. These methods are valid for any multivariate function and not specific to two-particle vertices.}

In the remainder of the paper, the so-called grid parameter $R$ will be of central importance, where the three-dimensional Matsubara frequency grid will consist of $2^{3R}$ grid points. Hence, this parameter governs the (exponential) number of points of the discretized grid and exponentially different length scales in the system and, thus, determines the length of the resulting QTT. This will become more clear in the following sections.

\subsection{Quantics tensor train representation}



\begin{figure}
    \centering
        \includegraphics{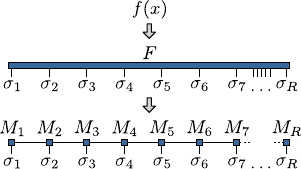}
        \label{fig:quantics-tensor-train-momentum}
    \caption{
        Quantics representation and quantics tensor train of a univariate function.
        } 
    \label{fig:quantics-tensor-train}
\end{figure}

Before discussing the three-dimensional Matsubara frequency case, let us introduce the QTT formalism for the one-dimensional case for educational purposes.
In the quantics representation, a discrete function $f(m)$ with $m \in \{0, \ldots, M-1\}$ on a one-dimensional grid with $M = 2^R$ grid points is instead seen as a $2\times 2\times ...\times 2$ ($R$ times) tensor \(F_{\sigma_1, \ldots, \sigma_R}\) (see Fig.~\ref{fig:quantics-tensor-train}), where each tensor index \(\sigma_1, \ldots, \sigma_R\) corresponds to a bit in a binary representation of \(m\):
\begin{equation}
    m = (\sigma_1 \sigma_2 \dots \sigma_R)_2 = \sum_{\ell=1}^{R} 2^{R-\ell} \sigma_{\ell},
    \quad
    \sigma_\ell \in \{0, 1\}
    \,, \label{eq:quanticsrep}
\end{equation}
with the discussed grid parameter $R$. Now, each bit corresponds to a distinct length scale of the system. The first bit \(\sigma_1\) represents the coarsest length scale which splits the system in halves, while the last bit $\sigma_R$ reflects the finest length scale. Hence, for continuous variables $m$ defined on a specific interval the grid parameter $R$ determines how exponentially dense the grid gets, while for discrete variables like Matsubara frequencies, $R$ determines how exponentially large the grid gets. In both cases, $R$ specifies the (exponential) number of grid points and a linear increase in $R$ corresponds to an exponential increase in the number of grid points.

This representation can be generalized to functions of \(N > 1\) variables by applying the binary representation to each variable separately. For instance, a function \(f(x, y, z)\) of three variables is represented as a tensor depending on \(3R\) binary indices
\begin{multline}
    F_{x_1, y_1, z_1, x_2, y_2, z_2, \ldots, x_R, y_R, z_R}
    =\\=
    f((x_1\ldots x_R)_2, (y_1\ldots y_R)_2, (z_1\ldots z_R)_2),
\end{multline}
and the \(L = 3R\) indices are relabeled to \(\sigma_1 = x_1, \sigma_2 = y_1, \sigma_3 = z_1, \sigma_4 = x_2, \ldots, \sigma_{L} = z_R\). Note that indices belonging to different variables are interleaved, which leads to an index ordering where all indices describing large length scales are grouped to the left, and all indices describing small length scales are grouped to the right. This tensor can then be factorized into a tensor train (TT),
also known as matrix product state (MPS), of the form
\begin{equation}
    F_{\sigma_1, \ldots, \sigma_L} \approx
    \prod_{\ell=1}^{L} M_{\ell}^{\sigma_\ell} =
    [M_1]^{\sigma_1}_{1\alpha_1} [M_2]^{\sigma_2}_{\alpha_1\alpha_2} \cdots [M_L]^{\sigma_L}_{\alpha_{L-1} 1}
    \,, \label{eq:TTdecomposition}
\end{equation}
with implied summation over repeated indices.
Each \(M_\ell\) is a three-leg tensor with local binary index \(\sigma_\ell\) and virtual indices \(\alpha_{\ell-1}, \alpha_{\ell}\), and we define the bond dimension \(D_\ell\) as the number of values that index \(\alpha_\ell\) is summed over. 
Hence, $M_\ell$ is a $D_{\ell-1} \times 2 \times D_\ell$ tensor. Generally, the bond dimensions \(D_\ell\) are truncated either at a fixed maximum bond dimension \(\Dmax\), or such that the factorization satisfies a specified error tolerance \(\epsilon\).
This truncated TT factorization can be performed using singular value decomposition (SVD), or using the tensor cross interpolation (TCI) algorithm (see the next Sec.~\ref{sec:tci}).

Overcoming the curse of dimensionality now depends on the maximum bond dimension $\Dmax=\textrm{max}_\ell(D_\ell)$, as the tensors \(M\) have \(\order(\Dmax^2 R)\) elements. The bond dimension required to reach a specified error tolerance \(\epsilon\) is strongly dependent on the structure of \(F\). If \(F\) is not compressible, e.g.~a random tensor, bond dimensions will grow exponentially with \(L\) as \(\Dmax\approx 2^{L/2}\) and the factorization will thus not result in an efficiency gain.
%
Fortunately, many functions in physics contain low-rank structure when factorized in their length scales. The interleaved representation groups bits corresponding to the same length scale, resulting in a highly compressed representation with small \(\Dmax\) \cite{Ritter2024,NunezFernandez2024}.


\subsection{Tensor Cross Interpolation (TCI)}\label{sec:tci}
The TCI-based factorization is performed by sampling a subset of the elements of the full tensor \(F\).
To be more specific, the TCI algorithm takes as input a tensor $F$ in the form of a function returning the value $F_{\sigma_1, \ldots, \sigma_L}$ at any given index $(\sigma_1, \ldots, \sigma_L)$~\cite{NunezFernandez2022,Ritter2024,NunezFernandez2024}.
The algorithm explores its structure by sampling in a deterministic way and constructs a low-rank approximation $\tilde F$ in the form of an MPS.
The algorithm increases the number of samples and the bond dimensions of the MPS adaptively,
until the estimated error $\epsilon$ in the maximum norm,
\begin{align}
    \epsilon = \frac{\|F-\tilde{F}\|_\infty}{\|F\|_\infty},
\label{eq:epsilon_TCI}
\end{align}
is below a specified tolerance.
Here, $\| \cdot \|_\infty$ denotes the maximum norm.

TCI is more efficient than the SVD-based factorization, especially when the full tensor does not fit into the available memory~\cite{NunezFernandez2022,Ritter2024,NunezFernandez2024}.
SVD-based factorization requires reading all elements of the tensor, leading to an exponential growth of the computation time in $R$.
In contrast, if the target tensor/function is low-rank, the computation time of the TCI-based factorization is linear in $R$, leading to an exponential speed-up over the SVD~\cite{Ritter2024}.
We refer the reader to Refs.~\onlinecite{ Ritter2024,NunezFernandez2024} for more technical details , e.g. for information on how the sampling points in the TCI algorithm are chosen. In the following computations TCI will only be used for compressing the initial input vertices and functions. On a more technical note, the \verb|crossinterpolate2| algorithm in the \verb|TensorCrossInterpolation.jl| library was used for compressing the objects. More details on this specific algorithm can be found in section 8.3.1 in Ref.~\onlinecite{NunezFernandez2024}.

\subsection{Matrix product operator (MPO)}\label{sec:mpo}
\begin{figure}
    \centering
    \includegraphics{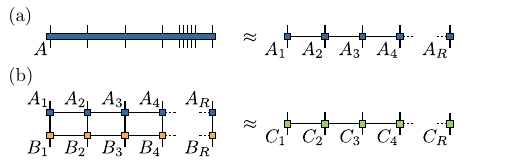}
    \caption{(a)~Decomposition of a tensor in MPO form. (b)~Contraction of two MPOs $A$ and $B$.
    }
    \label{fig:mpo}
\end{figure}

We use matrix product operators (MPOs) to perform operations on QTTs.
As illustrated in Fig.~\ref{fig:mpo}(a), an MPO of length $L$ has two physical legs on each tensor.
The MPO can be regarded as the factorization of a full tensor of order $2L$, or as a linear operator acting on an MPS of length $L$.

As we will see in later sections, many operations in QTT can be implemented as the contraction of two MPOs, illustrated in Fig.~\ref{fig:mpo}(b).
The exact contraction will result in an MPS of large bond dimension $D_\mathrm{A} D_\mathrm{B}$, where $D_\mathrm{A}$ and $D_\mathrm{B}$ are the bond dimensions of the two input MPOs, respectively.
Thus, the bond dimension of the resulting MPO must be truncated to some $\Dmax$.
The computational cost of a naive SVD-based contraction followed by truncation scales $\order(D_\mathrm{A}^3 D_\mathrm{B}^3)$.

In the present study, we will deal with two distinct cases: (a)~$D_\mathrm{A} = \order(1) \ll D_{B}$ (channel transformation), (b)~$D_\mathrm{A} = D_\mathrm{B} = \Dmax$ (Bethe--Salpeter equation).
For the former case (a), the naive approach is efficient enough, with scaling \(\order(D_{\mathrm{B}}^3 L)\).

However, a more efficient scheme is necessary for case~(b).
We use two algorithms: the ``fit algorithm'' fits a new MPOs to the MPO-MPO contraction~\cite{Verstraete2004}, and the ``zip-up algorithm'' combines contraction of core tensors with truncation of the bond dimensions \cite{Stoudenmire2010}.
We typically combine these, using the zip-up algorithm to generate an initial guess for the fit algorithm. If the resulting MPO is truncated to bond dimension $\Dmax = D_\mathrm{A} = D_\mathrm{B}$, the computational cost of both algorithms scales as \(\order(\Dmax^4 L)\). 
We want to emphasize that in the current implementation the combination of the two algorithms is still SVD- and not TCI-based. Thus, TCI is only used in the compression of the input functions and not in the MPO-MPO contractions. However, in future work we plan on using TCI for MPO-MPO contractions as well, since this is expected to be computationally more efficient.

\section{Parquet equations in QTT format}
\label{sec:parquet-in-qtt}

In order to evaluate the full set of parquet equations completely within the QTT representation, we need to (i) represent the vertex functions in the QTT form; and
(ii) decompose Eqs.~\eqref{eq:bse}-\eqref{eq:sde} into operations on QTTs.
 The latter can be implemented in a straightforward way by employing fundamental operations described in the previous section, namely MPO-MPO contractions.  In the following subsections, we describe the quantics  representation of vertex functions, check their compressibility to QTTs, and discuss the implementation of each step in solving the parquet equations. Two operations are particularly important: 
 \begin{itemize}
     \item [(1)] Affine transformations that are represented by an MPO with maximum bond dimension of $\order(1)$, needed in  the parquet equation Eq.~\eqref{eq:parquet-density} for frequency channel transformations of vertex functions (Sec.~\ref{sec:channel-transformation}); 
     \item[(2)]   Elementwise and matrix multiplications of two QTT vertex objects for solving the BSEs~\eqref{eq:bse} and SDE~\eqref{eq:sde}. In this case, auxiliary MPOs are introduced substituting the MPSs and then MPO-MPO contractions are applied (Sec.~\ref{sec:bse}).
\end{itemize}

%

\subsection{\label{sec:tci-compression}Quantics representation and compression of two-particle vertex functions}

\begin{figure}
    \centering
    \includegraphics{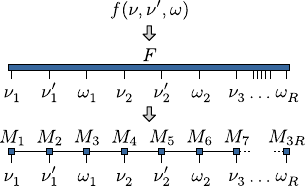}
    \caption{
        QTT representation for full vertex function.
        }
    \label{fig:qtt-vertex}
\end{figure}

{In this work, all functions represented in QTT format are functions of bosonic and fermionic frequencies, which are parameterized as $\nu = (2m - 2^R + 1)\pi/\beta$ and $\omega = (2m - 2^R)\pi/\beta$, respectively. The discrete index \(m \in \{0, \ldots, 2^R-1\}\) is then decomposed into quantics bits as in Eq.~\eqref{eq:quanticsrep}, and bits corresponding to different variables are then interleaved as illustrated in Fig.~\ref{fig:qtt-vertex}.}

\begin{figure}
    \centering 
    \includegraphics[width=\linewidth]{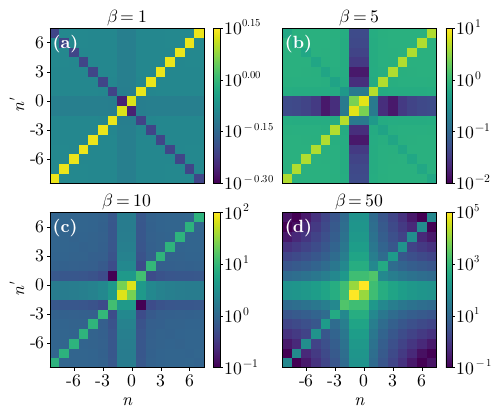}
    \caption{Absolute value of the full vertex in the density channel $F_d$ at $\omega=0, U=1$ for the $16$ innermost fermionic Matsubara frequencies
    $\nu^{(\prime)} = (2n^{(\prime)}+1)\pi / \beta$ for $\beta=1,5,10,50$.
    }
    \label{fig:tci-full-density}
\end{figure}

\begin{figure}
    \centering 
    \includegraphics[width=\linewidth]{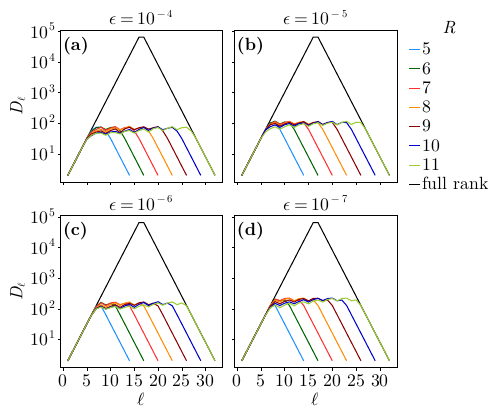}
    \caption{Bond dimension $D_\ell$ at bond $\ell$ for different tolerances set in the TCI construction of the QTT of $F_d$ in the interleaved representation for $\beta=U=1$. The different values of $R$ correspond to different grid sizes ($2^{3R}$ grid points). The black line indicates the exponentially growing bond dimension of the full rank QTT without any truncation for $R=11$.}
    \label{fig:tci-compression-tolerances}
\end{figure}

The first step in using the QTT framework for solving the parquet equations is to investigate the compressibility of vertex functions in the above representation. We use the full vertex in the density channel ($F_d$) of the Hubbard atom for numerical demonstration. The fermionic frequency dependence of $F_d$ at $\omega=0$ is shown for various temperatures in Fig.~\ref{fig:tci-full-density}.
Note, that in the Hubbard atom the results are not separately dependent on temperature and $U$, but only on their ratio $\beta U$, since there are no other energy scales \cite{Thunstroem2018}.
Before we focus on the scaling of bond dimension with temperature, let us first look at the bond dimension along the QTT at \(\beta U = 1\) for different grid sizes and tolerances set in TCI as shown in Fig.~\ref{fig:tci-compression-tolerances}. Moving inward from the first and last bonds, the bond dimension grows exponentially as $D_\ell = \min(2^\ell, 2^{\cL - \ell})$, which is the maximum bond dimension of an uncompressed factorization and represents maximum entanglement between these exponentially different length scales.
In between, the bond dimension then saturates at a maximum bond dimension \(\Dmax\), therefore indicating that the vertex structures are indeed compressible. The maximum bond dimension \(\Dmax\) is between \(80\) and \(400\), and increases with decreasing tolerance \(\epsilon\).
Importantly, \(\Dmax\) is nearly independent of the grid parameter \(R\), an exponential increase in the number of grid points ($\order(2^{R})$) can be achieved for linear cost ($\order(R)$) in runtime and memory.
These findings are consistent with earlier results on the compressibility of vertices in Ref.~\onlinecite{Shinaoka2023}.

For large temperatures, such as \(\beta U = 1\) in the above example, the dominating structures are the diagonal and anti-diagonal part of the vertex. For small temperatures, i.e.\ large \(\beta U\), the anti-diagonal vanishes and an additional cross structure appears for one of the Matsubara frequencies equal to $\pm \pi/\beta$. The source and physical meaning of different structures in two-particle vertices have been discussed in Ref.~\cite{Rohringer2012}. For large \(\beta U\), the vertex is large even in some places where both Matsubara frequency arguments are large, as can be seen in Fig.~\ref{fig:tci-full-density}(d). All in all, the structure of $F_d$ seems to be much simpler at large temperatures than at small ones, which leads us to expect an increase in bond dimension in QTT representation with increasing $\beta U$.

\begin{figure}
    \centering 
    \includegraphics[width=\linewidth]{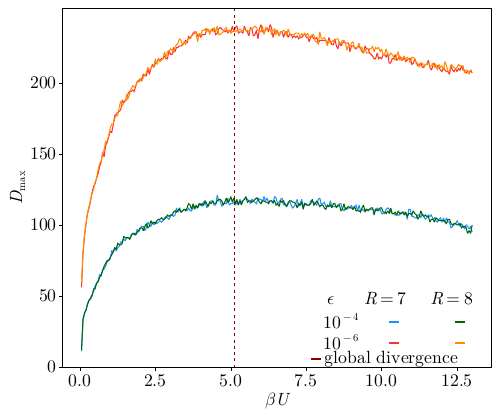}
    \caption{Maximum bond dimension $\Dmax$ of the QTT of $F_d$ for different grid sizes and various tolerances set in the TCI construction as a function of $\beta U$.}
    \label{fig:tci-maxbonddim-betaU}
\end{figure}

In Fig.~\ref{fig:tci-maxbonddim-betaU}, we show $\Dmax$ of the QTT representation of $F_d$ for various values of $\beta U$, different tolerances set in TCI, and various grid sizes. As expected from earlier considerations, the maximum bond dimension is quite low at high temperatures and steeply grows with $\beta U$ until $\beta U \approx 5.1$, where it reaches a maximum. Thereafter, the bond dimension decreases again, which does not conform with our expectations. The maximum can be related to the first global divergence of the irreducible vertex $\Gamma_d$~\cite{Thunstroem2018, Schaefer2013, Schaefer2016, Chalupa2021, Rohshap2023, Pelz2023}, which occurs at  $\beta U = 5.13715$ and is indicated by a dashed line in the plot.
Although the irreducible vertex $\Gamma_d$ diverges at this value of  $\beta U$, there is no phase transition connected with the divergence and $F_d$ remains finite.
The presence of a maximum in $\Dmax$ precisely at the first global divergence of $\Gamma_d$ is very interesting from two different perspectives. Firstly, we see that $\Dmax$ stops growing with $\beta U$ and, thus, calculations for temperatures ranging from very low to very high are manageable within the QTT framework in this case. Secondly, although the full vertex $F_d$ does not contain singularities, we see fingerprints of this first global divergence of $\Gamma_d$ in the amount of length scale entanglement in the system represented by the maximum bond dimension. A deeper investigation of this behavior is left for future work.

The vertex functions in other channels show a similar scaling behavior as $F_d$ (not shown here), and hence it can be concluded that vertex functions of the Hubbard atom are nicely compressible with a maximum bond dimension of around $100$. Together with the logarithmic scaling in grid size, this indicates that QTCI can indeed overcome memory and computational bottlenecks when dealing with two-particle vertices. 

The Hubbard atom is quite an extreme limit of the Hubbard model and one could think that the high compressibility is related to 
structures present in the vertices that are limited to this atomic limit. However, this is not the case:  prominent structures in the frequency dependence of vertices are well-known to be present in the SIAM (see Sec.~\ref{sec:results} below for an example), as well as in the local approximation of the Hubbard model (in dynamical mean-field theory)~\cite{Rohringer18, Chalupa2021}. These structures arise specifically from certain types of diagrams~\cite{Rohringer18, Chalupa2021} and also from insertions from one scattering channel to another through the parquet equation~\eqref{eq:parquet}. For models whose vertices depend on momentum and/or orbital indices, the structures may become more complicated. Nevertheless, their origin is well
understood and we expect the vertices to generically have structures for a large range of parameters and models. As long as these structures show some kind of scale separation, they will be QTT compressible, hopefully with still manageable bond dimensions.




\subsection{\label{sec:channel-transformation}Channel transformations}
\begin{figure}
    \centering 
    \includegraphics{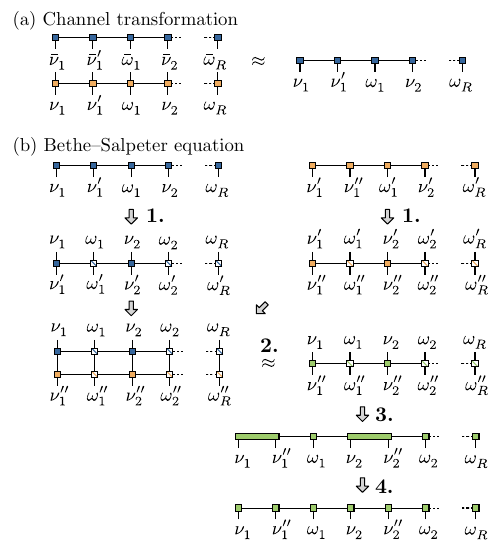}
    \caption{QTT implementation of the Bethe--Salpeter equations as tensor networks.
    (a) Channel transformation of a vertex in QTT form (blue) using an affine transform MPO (orange). This is described in more detail in Appendices~\ref{app:mpo-affine} and \ref{app:channel-transformation}.
    (b) The Bethe--Salpeter equations are evaluated from QTT vertices and are implemented using multiple MPO-MPO contractions (see text). The contraction itself is done in 4 steps: 1.~Both QTT to be contracted are converted to MPOs. 2.~The MPOs are contracted to a single MPO. 3.~The duplicate \(\omega_\ell''\) legs are removed.
    4.~The tensors with \(\nu_\ell\) and \(\nu''\) legs are factorized between their local legs, reaching the original QTT form.
    }
    \label{fig:diagram-parquet-qtt}
\end{figure} 
Each two-particle reducible vertex \(\Phi\) is parameterized as a function of two fermionic frequencies \(\nu, \nu'\) and one bosonic frequency \(\omega\). Before performing the sum in the parquet equation~\eqref{eq:parquet-density}, it is necessary to perform transformations such as \(\Phi_d^{\nu \nu' \omega} \rightarrow \Phi_d^{\nu (\nu+\omega) (\nu' - \nu)}\) to translate between the frequency parametrizations corresponding to different channels \cite{Bickers04}, as discussed in Sec.~\ref{sec:parquet} and App.~\ref{app:fermionic-frequencies}.
In QTT format, transforming the function arguments is a non-trivial task, since each argument is split into bits across different tensor indices. Affine transformations such as the channel transformations needed here can be expressed as MPOs with small bond dimensions,
as described in App.~\ref{app:mpo-affine}. The QTT implementation of channel transformations is introduced explicitly in App.~\ref{app:channel-transformation}.
As illustrated in Fig.~\ref{fig:diagram-parquet-qtt}(a), these MPOs are then applied to vertex QTTs, followed by truncation of the QTT to the specified bond dimension.

\begin{figure}
    \centering 
    \includegraphics[width=\linewidth]{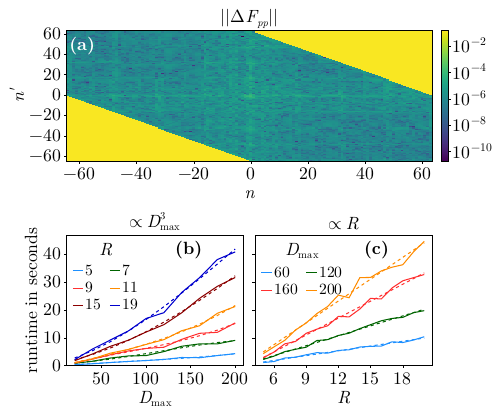}
    \caption{(a) Absolute normalized error $||\Delta F_{\pp}|| := |F_{\pp, \mathrm{trafo}} - F_{\pp, \mathrm{exact}}|/||F_{\pp, \mathrm{exact}}||_{\infty}$ of the \ph\ to \pp\ channel transformation of $F$ at $\omega=0$ in the fermionic Matsubara frequency plane for a $D_{max} = 100$ and $R=7$, $\beta = U=1$, $\epsilon=10^{-8}$.
    \mbox{(b)--(c)} Runtime of the \ph\ to \pp\ channel transformation of $F$ for various maximum bond dimensions and grid size parameters $R$ with $\beta=U=1$. The dashed lines indicate (b) the cubic runtime increase with $\Dmax$, and (c) the linear increase by increasing $R$, which corresponds to an exponential increase in the number of grid points.}
    \label{fig:channel-transformation-error}
\end{figure}

In this process, there are two distinct sources of error: the finite frequency box and the QTT truncation.
As an example, Fig.~\ref{fig:channel-transformation-error}(a) shows the absolute normalized error ($||\Delta F_{\pp}|| := |F_{\pp, \mathrm{trafo}} - F_{\pp, \mathrm{exact}}|/||F_{\pp, \mathrm{exact}}||_{\infty}$) of a transformation of the full vertex $F$ in the \ph\ channel parametrization  to the \pp\ channel parametrization (denoted as $F_{\pp}$) in the $\nu,\nu'$-plane.
In two triangular regions, the upper right and lower left corner, errors are large due to the finite size of the frequency box, and are not caused by the QTT compression. These points correspond to grid points outside of the original frequency box that were transformed into the box by the \ph\ to \pp\ transformation. The missing data there can be either replaced by zeros (which corresponds to open boundary conditions of the affine transformation, see App.~\ref{app:mpo-affine}) or by values extrapolated from another part of the frequency box (e.g. by using periodic boundary conditions for the affine transformation, as in Ref.~\cite[footnote 14]{YangPRE09}). We checked that for the examples shown in this paper the average difference in error between the two options is small. We used periodic boundary conditions for all results presented in the manuscript. For the half-filled Hubbard atom this leads to better representation of missing values on the diagonal due to high symmetry of the full vertex in this case. In general, the choice of boundary conditions can be adapted to the problem at hand. Since the error 
can easily be reduced and shifted to higher Matsubara frequencies by increasing the size of the frequency box exponentially through increasing $R$, we do not expect the choice of boundary conditions to have significant effect on the error.

In the remaining diagonal region in between the two yellow corners in Fig.~\ref{fig:channel-transformation-error}(a), the error is entirely due to the tensor train approximation. At a bond dimension of \(\Dmax = 100\), the error is smaller than the error tolerance of $\epsilon = 10^{-5}$ everywhere, meaning that the crossing symmetry is also fulfilled up to this error.

Since this operation consists of a single MPO-MPS contraction, it is expected to scale as \(\order(\Dmax^3 L) = \order(\Dmax^3 R)\) provided that the bond dimension is independent of \(R\). We verify this explicitly in Figs.~\ref{fig:channel-transformation-error}(b-c). Compared to increasing resolution, decreasing error tolerances and thus increasing $\Dmax$ is more expensive.

\subsection{Parquet equation}
The parquet equation \eqref{eq:parquet} with its frequency shifts as in \eqref{eq:parquet-density} can be solved entirely in QTT by first converting all the vertex functions \(\Lambda, \Phi\) to the required channel as described in the previous section, then performing their summation and subtraction as shown in Ref.~\cite[Sec.~4.7]{NunezFernandez2024}.
The resultant QTTs for $F$ (and subsequently \(\Gamma\) obtained from \eqref{eq:phi}) are then compressed to a maximum bond dimension \(\Dmax\) in \(\order(\Dmax^3 R)\) computation time. Hence, the parquet equation has the same $\order(\Dmax^3 R)$ computational cost as channel transformations. Further investigation of error and runtime scaling can be found in App. \ref{app:qtt-parquet}.

\subsection{Bethe--Salpeter equation}
\label{sec:bse}
The costliest part of the iterative parquet scheme are the Bethe--Salpeter equations \eqref{eq:bse}, where two infinite Matsubara sums have to be performed. These can be implemented as a sequence of matrix multiplications as $(\Gamma \chi_0) F$, where $\chi_0$ is treated as a vertex object with two fermionic frequency axes and one bosonic frequency axis.
At each multiplication step, we have to compute the product of two vertex functions $A$ and $B$ as
\begin{align}
    C^{\nu\nu''\omega} &= \sum_{\nu'} A^{\nu \nu' \omega} B^{\nu' \nu'' \omega}.
\end{align}
To express this summation as matrix multiplication, we introduce dummy indices \(\omega'\) and \(\omega''\), such that
\begin{subequations}
\begin{align}
    C^{\nu \nu'' \omega} &= \sum_{\nu'\omega'} \tilde A^{\nu \omega}_{\nu' \omega'} \tilde B^{\nu' \omega'}_{\nu'' \omega''} \Big |_{\omega = \omega''}, \label{eq:matmul:dummy} \\
    \tilde A^{\nu \omega}_{\nu' \omega'} &:= A^{\nu \nu' \omega} \delta_{\omega, \omega'}, \\
    \tilde B^{\nu' \omega'}_{\nu'' \omega''} &:= B^{\nu' \nu'' \omega'} \delta_{\omega', \omega''},
\end{align}
\end{subequations}
where $|_{\omega = \omega''}$ denotes the restriction of the result to $\omega = \omega''$.
Note that Eq.~\eqref{eq:matmul:dummy} has the structure of a matrix multiplication in the combined index \((\nu', \omega')\).
Thus, this equation can be evaluated in QTT format through standard MPO-MPO contraction, as is illustrated in Fig.~\ref{fig:diagram-parquet-qtt}(b). 
For numerical computations, and, thus, also for the QTT approach, the infinite Matsubara sums have to be truncated. The number of Matsubara frequencies taken into account is governed by the chosen Matsubara grid,
containing $2^R$ points in each direction.
Introducing dummy indices has a runtime and memory cost of \(\order(\Dmax^2 R)\), which is much smaller than the cost of other steps in the algorithm.

After each MPO-MPO contraction, the bond dimensions are truncated to \(\Dmax\).
If all MPOs are truncated to \(\Dmax\), the computational cost of each contraction is expected to scale as \(\order(\Dmax^4 R)\) as described in Sec.~\ref{sec:mpo}, which is more expensive than the channel transformation for large \(\Dmax\).

\begin{figure}
    \centering 
    \includegraphics[width=\linewidth]{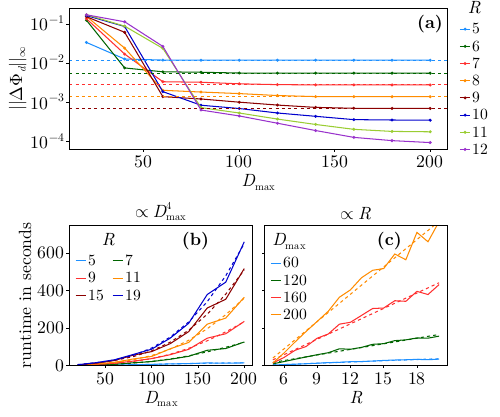}
    \caption{(a) Dependence on \(\Dmax\) of the maximum absolute normalized error $||\Delta \Phi_d||_{\infty} := ||\Phi_{d,\mathrm{BSE}} - \Phi_{d,\mathrm{exact}}||_{\infty}/||\Phi_{d,\mathrm{exact}}||_{\infty}$ of the BSE with QTCI for the reducible vertex $\Phi_d$ compared to the exact values, for various grid sizes, with $U=\beta=1$ and $\epsilon= 10^{-10}$. The dashed lines indicate the results from the dense grid calculation without QTCI.
    (b)--(c) Runtime of a single evaluation of the BSE for $\Phi_d$ for various maximum bond dimensions and grid size parameters $R$ with $\beta=U=1$. The dashed lines indicate (b) the quartic runtime increase with $\Dmax$, and (c) the linear increase by increasing $R$, which corresponds to an exponential increase in the number of grid points.}
    \label{fig:bse-error}
\end{figure}

We now conduct numerical tests to verify the accuracy of the operation and the scaling of the computational cost.
Figure~\ref{fig:bse-error}(a) shows the dependence on \(\Dmax\) of the maximum absolute normalized error ($||\Delta \Phi_d||_{\infty} := ||\Phi_{d,\mathrm{BSE}} - \Phi_{d,\mathrm{exact}}||_{\infty}/||\Phi_{d,\mathrm{exact}}||_{\infty}$) [cf. Eq.~\eqref{eq:epsilon_TCI}] of the QTT implementation of the BSE in the density channel for various grid sizes. The dashed lines indicate the results applying these matrix multiplications for the full numerical data and without the compressed QTTs. The error of these ``dense grid calculations'' is due to the finite size of the grid and, thus, caused by the truncated Matsubara sum. The results can be improved by increasing the grid size.

For the dense grid calculations, the improved results come at high cost since increasing the grid parameter $R$ leads to an exponential increase in memory and computational cost. By contrast, when using QTTs the memory and computational cost only increase linearly with $R$, which can be observed in Fig.~\ref{fig:bse-error}(c), where the linear dependence of the runtime of the BSE on $R$ is shown for different maximum bond dimensions. 

Moreover, it can be observed that the QTT BSE errors converge to the box results. Interestingly, for larger grid sizes, slightly larger $\Dmax$ are needed to reach the same error level as in the case of smaller boxes. However, a maximum normalized error $ <10^{-3}$ can be easily reached using QTTs for a still reasonable bond dimension of 200. Without the use of QTTs this would correspond to calculations with objects of $8 \times 2^{3\times 12} \simeq 5.5\times 10^{11}$ bytes, for which multiple nodes on a cluster would need to be occupied. With the current SVD-based QTT matrix multiplication implementation the BSE operation only takes about 400 seconds on a single 512 GB node (equipped with two AMD EPYC 7713 processors) without parallelization. 
 The bottleneck of performing the BSE is, as expected, the quartic dependence on $\Dmax$, which can be observed in Fig.~\ref{fig:bse-error}(b).
Overall we numerically verified $\order(R \Dmax^4)$ computational cost for the BSE using QTTs.
Since the error of the result of the BSE is bound by the grid size, QTTs provide an efficient way to overcome this bottleneck.

\subsection{Schwinger--Dyson equation}
Similarly to the BSE evaluation, two infinite Matsubara sums have to be performed in the SDE in Eq.~\eqref{eq:sde}. Hence, qualitatively similar scaling with \(\Dmax\) and \(R\) to the BSE case are expected. This is indeed the case. The results are presented in Fig.~\ref{fig:sde-error} in Appendix~\ref{app:qtt-sde}.

\subsection{Initialization and update strategy}
Let us briefly discuss how the iterative parquet calculations are performed in practice. 
We start by compressing the initial inputs $G_0$ and the exact fully irreducible vertices $\Lambda_r$ with TCI, where in the Hubbard atom case the exact $\Lambda_r$ is taken as input and in the SIAM case we make use of the parquet approximation ($\Lambda_d = U, \Lambda_m = -U, \Lambda_s = 2U, \Lambda_t = 0$). We then set $\Gamma_r = \Lambda_r, F_d = U,  F_m = -U, F_s = 2U,  F_t = \Lambda_t, G=G_0, \Phi_r=0$, where the QTT representation for the input full vertices $F_r =U$ can easily be obtained with or without TCI since this can be represented by a QTT of bond dimension one.
Then, the right sides of the four BSEs are computed 
using (SVD-based) MPO-MPO contractions that were discussed in Sec. \ref{sec:mpo}. The output of the BSEs is then linearly mixed with the QTTs of the input $\Phi_r$ resulting in updated QTT approximations of the reducible vertices $\Phi_r$. These updated $\Phi_r$ QTTs are then used as input in the parquet equations; as output, the QTT representations of $\Gamma_r$ and $F_r$ are updated. As a last step the QTT representation of the self-energy is updated by solving the SDE in QTT format. The updated QTT vertices and self-energy are then used as new input in the next iteration step of the iterative parquet scheme.

\section{Results of self-consistent calculations}

\label{sec:results}

With all the components from the previous section in place, the parquet equations can now be iteratively solved within the developed two-particle QTT framework for our two test cases: the Hubbard model in the atomic limit and the single-impurity Anderson model.  



\subsection{Hubbard atom}
The atomic limit of the Hubbard model gives rise to rich two-particle correlations giving insight into strong-copupling  limit of the Hubbard model \cite{Rohringer2012, Schaefer2016, Thunstroem2018}.
This atomic limit offers considerable simplification, as the vertex functions are analytically known and become independent of momentum \cite{Thunstroem2018}. Therefore, the Hubbard atom serves as an ideal first test case for exploring two-particle properties in strongly correlated electron systems using the QTT framework.

Following the iterative parquet scheme outlined in Fig.~\ref{fig:iterative-parquet}, we start from the exact fully irreducible vertices $\Lambda_r$, set $\Gamma_r = \Lambda_r, F_d = U,  F_m = -U, F_s = 2U,  F_t = \Lambda_t, G=G_0$ and use TCI to efficiently compress the data into QTTs. We then iterate the four BSEs, the  parquet equation and the SDE in QTT format by means of the discussed MPO operations, which leads to quick convergence of the results for $\beta=U=1$.

\begin{figure}
    \centering 
    \includegraphics[width=\linewidth]{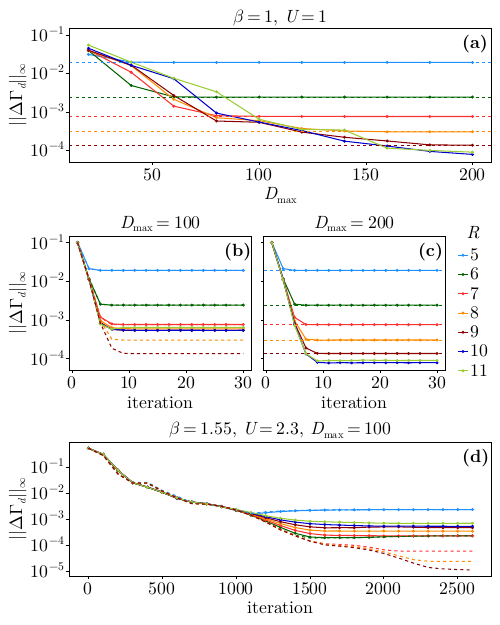}
    \caption{(a) Maximum absolute normalized error $||\Delta \Gamma_d||_{\infty} := ||\Gamma_{d,\text{iterative-parquet}} - \Gamma_{d,\text{exact}}||_{\infty}/||\Gamma_{d,\text{exact}}||_{\infty}$ of the iterative parquet scheme with QTCI for the irreducible vertex $\Gamma_d$ compared to the exact values for various grid sizes ($2^{3R}$ grid points) in the case of $U=\beta=1$, plotted (a) as a function of $\Dmax$ after 30 iterations, and as a function of iteration number for (b) $\Dmax=100$ and (c) $\Dmax=200$. Dashed lines indicate the results from the dense grid calculation without QTCI. (d) The maximum absolute normalized error of $\Gamma_d$, shown close to the first divergence, for $\beta=1.55, U=2.3$ and $\Dmax=100$, up to a very large number of 2600 iterations.
    }
    \label{fig:iterative-parquet-sde-error}
\end{figure}

Figure~\ref{fig:iterative-parquet-sde-error}(a) shows the maximum absolute normalized error in $\Gamma_d$ compared to the exact result after 30 iterations of the iterative parquet cycle with a set tolerance of $10^{-10}$ in the initial TCI. We can now disentangle the two sources of error\textemdash{}the finite size of the discrete frequency grid, corresponding to the error in the respective dense grid calculations indicated by dashed lines, and the QTT approximation. The error due to the QTT approximation for a specified maximum bond dimension can be identified as the difference between the QTT and the dense grid results. Remarkably, the QTT errors quickly converge towards the dense grid results, where larger bond dimensions are needed to reach lower errors for larger grids. E.g. in the case of $R=9$ already with $\Dmax=180$, the error from the QTT approximation de facto vanishes leading to the same result as the dense grid calculation. However, the difference is that in the dense grid calculations objects containing $2^{3R} \simeq 1.34 \times 10^{8}$ data points need to be stored, while the QTTs stored for these parameters consist only of $\sim 8.5 \times 10^{5}$ elements leading to a compression ratio of $\order(10^2)$, remarkably, without any loss of accuracy.

If we allow for a small loss of accuracy, even more impressive compression ratios can be achieved, while at the same time reaching very low errors. For instance, it can be observed that already at a bond dimension of 100 maximum normalized errors $< 10^{-3}$ can be achieved. In the case of $R=11$ this corresponds to a compression ratio of  $\order(10^{4})$ leading only to a tiny fraction of the required memory occupation in comparison to the dense grid calculations. Thus, we observe that the outlined framework can be used to efficiently solve the parquet equations in the compressed QTT format. 

In Figs.~\ref{fig:iterative-parquet-sde-error}(b) and (c), we show the maximum normalized errors for various grid sizes with respect to the iteration. At $\Dmax=100$ (b) the dense grid results (dashed lines) are only up to $R=7$ exactly reproduced, while the errors of larger grid calculations level off below $10^{-3}$ in the vicinity of each other. Furthermore, we see that for larger grids slightly larger maximum bond dimensions are required to obtain the same level of error. In comparison, at $\Dmax=200$ (c) the QTT calculations converge towards the resulting dense grid errors also up to $R=9$. 

Next, we show that computations can also be performed for a more challenging case with the parameters $\beta=1.55, U=2.3$, which are chosen in the same way as in Ref. \citep{Wallerberger2021}. This case is  interesting since $\beta U = 3.565$ is very close to the point, where the irreducible vertex in the density channel diverges ($\beta U \approx 3.628$)~\cite{Kozik2015, Gunnarsson2017, Essl2024, Reitner2024, Reitner2024b, Essl2025}. We show the results of these calculations in Fig.~\ref{fig:iterative-parquet-sde-error}(d) for a maximum bond dimension $\Dmax=100$, where the maximum normalized error of $\Gamma_d$ is shown with respect to the iteration. Since the parameters are very close to the first divergence line, we use a small mixing parameter $\alpha=0.01$, leading to convergence only after 2600 iterations. In accordance with (b), we observe that in the case of $R \leq 6$ the error plateaus at decreasing levels for increasing $R$, which is due to the increased box size reflecting the results from the dense grid calculation. However, for $R \geq 7$ the situation changes, where the leveled off errors are closer together. This means that the error is now governed by the QTT approximation (i.e.\ by 
$\Dmax$) and not by the finite box size anymore. Still, it can be seen that already a maximum bond dimension of 100 is sufficient to reach absolute normalized errors $< 10^{-3}$ for larger values of $R$, similar as in (b). 

The above analysis demonstrates that during BSE iterations, errors due to the QTT representation do not accumulate  -- if they did, the solid lines in Figs.~\ref{fig:iterative-parquet-sde-error}(b,c) would slope upwards with increasing iteration number.
Instead, the error saturates, at a value governed by the maximum of the error due to the Matsubara sum truncation and the initial QTT approximation. The accuracy of the results can be improved systematically by increasing the maximum bond dimension. In principle, this can also be done during the course of the BSE iterations if these generate  structures of increasing complexity, but that was not necessary for the calculations presented here.

\begin{figure}
    \centering 
    \includegraphics[width=\linewidth]{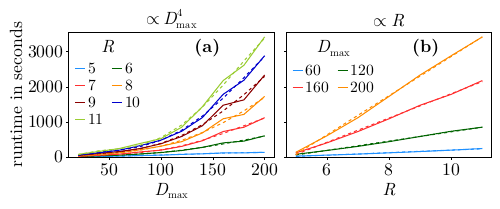}
    \caption{Runtime of a single iteration of the iterative parquet scheme for various maximum bond dimensions and grid size parameters $R$ in the case of $\beta=U=1$. The dashed lines indicate (a) the quartic runtime scaling with $\Dmax$ (governed by the BSEs), and (b) linear increase with $R$.}
    \label{fig:timing-iterative-parquet}
\end{figure}

Our calculations were performed on a single 512 GB node on a cluster without parallelization, where in Fig.~\ref{fig:timing-iterative-parquet}, the runtime of a single iteration of the iterative parquet scheme is shown. For the dense grid calculations performing the same iterative parquet cycle was possible only up to $R=9$ due to the exponentially increasing memory demand. In contrast, using the QTT approach, calculations for $R=11$ were easily carried out on this single node without parallelization. This demonstrates the advantage of QTTs, where memory occupation and operations scale logarithmically with increasing resolution (see Fig.~\ref{fig:timing-iterative-parquet}(b)), in contrast to the rapid growth in memory and computational costs encountered in standard methods. This allows for efficient computation on large grids, providing a significant advantage over dense grid implementations.

\subsection{Single-impurity Anderson model}
After solving the parquet equations for the simplified Hubbard atom case, we extend the QTT framework to the more complex SIAM, where a Hubbard atom-like interacting site is coupled to a bath of non-interacting electrons. 
Using the parquet approximation in which the fully irreducible vertex is approximated by the bare interaction ($\Lambda_d = U, \Lambda_m = -U, \Lambda_s = 2U, \Lambda_t = 0$), we iteratively evaluate the four BSEs~\eqref{eq:bse}, the parquet equation~\eqref{eq:parquet} and the SDE~\eqref{eq:sde} with QTTs. Starting from $\Gamma_r = \Lambda_r, F_r = \Lambda_r, G = G_0$, we decompose the relevant functions into QTTs using TCI and then make use of the discussed MPO operations in order to iteratively solve the parquet equations. To ensure full convergence, we perform 60 iterations with a linear mixing $\alpha = 0.4$.  The results presented below were obtained for $\beta = 10$, $U=1$, $V=2$, $D=10$ and half-filling, i.e.\ with $\varepsilon_d=-U/2$. For these parameters, the SIAM is in the weakly correlated regime, where the parquet approximation still holds. We compare our results with reference data obtained for the parquet approximation with the state-of-the-art parquet equations implementation on large equidistant frequency grids of Ref.~\onlinecite{Krien2022} using the single- and multi-boson exchange formulation~\cite{Krien2020b}.
\begin{figure}[t]
    \centering 
    \includegraphics[width=\linewidth]{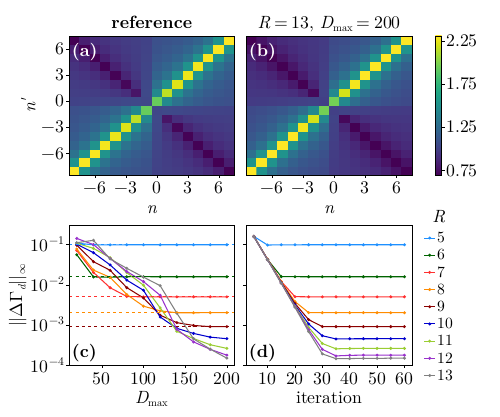}
    \caption{Irreducible vertex $\Gamma_d$ of the reference data (a) compared to the iterative parquet approximation calculations for $R=13, \Dmax = 200$ (b) for $\beta=10, U = 1, V=2, D=10, \epsilon=10^{-6}$ and a mixing of $0.4$.
    (c)--(d) The maximum normalized error of $\Gamma_d$ ($||\Delta \Gamma_d||_{\infty} := ||\Gamma_{d,\mathrm{iterative-parquet}} - \Gamma_{d,\mathrm{ref}}||_{\infty}/||\Gamma_{d,\mathrm{ref}}||_{\infty}$) with respect to the reference data is shown (c) as a function of the maximum bond dimension $\Dmax$ and (d) dependent on the iteration for $\Dmax =200$, where dashed lines indicate the obtained errors from dense grid calculations. }
    \label{fig:AIM-error}
\end{figure} 

Fig.~\ref{fig:AIM-error}(b) shows the irreducible vertex $\Gamma_d$ at $\omega=0$ calculated for $R=13$ and a maximum bond dimension $\Dmax=200$, which is in good agreement with reference data in (a). In (c), we show the maximum normalized error of $\Gamma_d$ obtained from the QTT calculations  in comparison to the reference data depending on the maximum bond dimension for various grid sizes. In agreement with the results for the Hubbard atom, it can be observed that the errors of the QTT calculations converge towards the results of the dense grid calculations, which are indicated by dashed lines. Like in the Hubbard atom case, we exactly reproduce the dense grid results at $R=9$ and $\Dmax = 180$ leading to a $\order(10^{2})$ compression ratio de facto without any loss in accuracy due to the QTT approximation. Moreover, since these calculations were performed up to $R=13$, very large compression ratios of $\order(10^{5})$ can be reached, while obtaining a maximum normalized error $< 10^{-3}$. In (d), the maximum normalized error with respect to the iteration for $\Dmax=200$ can be observed. We show that for larger grids more iterations are needed to converge due to approaching smaller errors.

Finally, let us mention that the performed calculations for $R=13$ would correspond to dense grid calculations with multiple objects of the size of $8 \times 2^{3 \times 13} \simeq 4.4 \times 10^{12}$ bytes. Instead of the necessity of engaging multiple nodes and making use of parallelization for the dense grid calculations, using the QTT framework it was possible to perform these calculations on a single node without parallelization. Applying this approach allowed us to obtain maximum normalized errors $< 10^{-3}$, while using only a tiny fraction of the memory required for the corresponding dense grid computations. This shows the computational advantage of the QTT approach.


\subsection{Technical limitations}
In our calculations, there are two main sources of error: (1) the finite size of the discrete Matsubara frequency grid, and (2) the QTT approximation, which is governed by the maximum bond dimension. The finite grid size determines how accurately we can evaluate the BSEs and SDE, as it controls the truncation of the infinite Matsubara sums. On the other hand, the QTT approximation dictates how accurately the data can be represented.

The combined QTT and TCI approach exhibits logarithmic scaling in both memory and computational costs relative to the grid size, enabling the potential to handle very large grids (e.g., $R = 20$).
While this is theoretically feasible, our explicit calculations for the Hubbard atom show that for such large values of $R$, the error increases significantly compared to the smaller grid sizes used in this study.
This issue is not inherent to the method itself but arises due to the current implementation of MPO-MPO contractions, which relies on bond dimension truncation via SVD.
The SVD truncation suffers from a loss of accuracy, such as round-off errors, because the Frobenius norm of the vertex functions diverges at large $R$ due to a constant term in the frequency domain.
This limitation can be addressed in future work by switching to a CI-based truncation approach~\cite{NunezFernandez2024}. For more details on this technical aspect, we refer readers to Appendix~\ref{app:technical-limitations}. Alternatively, the vertex asymptotics can be explicitly removed from parquet equations as in Ref.~\onlinecite{Krien2020b} or~\onlinecite{Wentzell2020}.

Finally, it is important to note that the maximum bond dimension directly governs the accuracy of the QTT approximation, as it reflects how compressible the data are. Large bond dimensions can significantly increase computational costs, making them the primary bottleneck for scaling up the calculations.

\section{Conclusion and outlook}
\label{sec:conclusions}

This work represents a large step forward in solving many-body problems with quantum field theory methods in QTT representations. The chosen example, the self-consistent solution of parquet equations, is a challenging one, requiring both efficiency in constructing the QTT representation of two-particle vertices and in evaluation of matrix multiplications and variable shifts within this representation. At the same time, the parquet equations for the simplest model, the Hubbard atom, can be solved analytically, allowing for careful benchmarking and assessment of the performance at each step of the solution separately. In this paper we have numerically shown that the QTT representation of the vertex frequency dependence is suitable for solving the parquet equations and that together with TCI it leads to only logarithmic scaling in the grid size and with fourth power in the maximum bond dimension. For the two examples of Hubbard atom and SIAM we observed that the bond dimension of $\sim$100-200 is enough to obtain the solution with high accuracy. For the case of Hubbard atom we see a saturation (or even decrease) of the bond dimension with increasing the inverse temperature $\beta$. We also expect a saturation or only slow growth of the bond dimension with $\beta$ in more general cases, as conjectured in Ref.~\onlinecite{takahashi2024compactnessquanticstensortrain}.

The naive iteration of the parquet equations is difficult to converge in some regimes, cf. Sec.~\ref{sec:iterparquet}. While this problem is almost orthogonal to questions of representation and compression of the vertices, QTTs offer potential synergies: in particular, the greatly reduced size of QTT vertices may enable a solver to keep a ``convergence history'' and use non-linear mixing schemes, which have been shown to stabilize convergence in Hartree-Fock~\cite{Pulay80}, or quasi-Newton solvers, which have been able to access previously hidden solutions in self-consistent diagrammatic theories~\cite{Strand2011}.

Although for explicit testing we have chosen models having no other degrees of freedom than frequencies (no momentum or orbital dependence), the dissection of the parquet equations solver into operations on QTTs\textemdash{}TCI compression and MPO-MPO contractions\textemdash{}is general and the extension to lattice models is straightforward. All results presented here were obtained on a single core with 512 GB memory and the grid sizes in each frequency variable were up to $2^{20}$. This high compressibility of the frequency dependence of vertex functions can in the future be exploited (i) to solve parquet equations for lattice systems with high momentum resolution and orbital degrees of freedom, needed to address material properties (see App.~\ref{app:extensions} for a brief discussion of how to deal with such additional degrees of freedom); 
(ii) to apply QTCI to other vertex based methods, such as the
functional renormalization group (fRG)~\cite{Hille2020,Niggemann2021, Bippus2024}; 
ladder extensions~\cite{Rohringer18} or fRG extensions \cite{Vilardi2019}
of dynamical mean-field theory (possibly using as input results for the local vertex obtained using the numerical renormalization group \cite{Kugler2021,Lee2021,Lihm2024}); embedded multi-boson exchange methods~\cite{Kiese2024b}; or the Migdal-Eliashberg theory in \textit{ab initio} calculations~\cite{Schrodi2020, Sano2016}.

On a more general note, two-particle objects are central in many more applications involving interacting electrons: in particular, the two-electron integrals $[ij|kl]$ are central to quantum chemistry, whereas the renormalized interaction $W_{ijkl}$ is one of the main ingredients of $GW$ \cite{Szabo1996}. Neither of these objects have the intricate three-frequency structure of the vertex $F$, however, they do depend on four orbital (spatial) indices. Handling this dependence is usually the main computational bottleneck in self-consistent field computations, even with sophisticated mitigation techniques~\cite{Dunlap2000}.  The present study offers a blueprint for applying QTTs to these methods and is a promising topic for future study.


\begin{acknowledgments}
This work was funded in part  by the Austrian Science Fund (FWF) Projects No.~P~36332 (Grant DOI 10.55776/P36332) and V~1018 (Grant DOI 10.55776/V1018). For open access purposes, the authors have applied a CC BY public copyright license to any author-accepted manuscript version arising from this submission. This work was funded in part by the Deutsche Forschungsgemeinschaft under Germany’s Excellence Strategy EXC-2111 (Project No.~390814868). M.K.R.\ and J.v.D.\ acknowledge support from DFG grant LE 3883/2-2.
This work is part of the Munich Quantum Valley, supported by the Bavarian state government with funds from the Hightech Agenda Bayern Plus. H.S.\ was supported by JSPS KAKENHI Grants No.~21H01041, No.~21H01003, No.~22KK0226, and No.~23H03817 as well as JST FOREST Grant No.~JPMJFR2232 and JST PRESTO Grant No.~JPMJPR2012, Japan.
Calculations have been partly performed on the Vienna Scientific Cluster (VSC) using the \verb|Quantics.jl| and \verb|TensorCrossInterpolation.jl| libraries. The raw data for the figures reported are available at Ref. \onlinecite{Rohshap2025Data}.
\end{acknowledgments}

\appendix

\begin{widetext}
\section{Parquet equation and frequency conventions \label{app:fermionic-frequencies}}
The parquet equation \eqref{eq:parquet} gives the full vertex $F$ as a simple sum of the fully irreducible vertex $\Lambda$ and vertices reducible in the \ph, \pp, and \phbar\ channels. The reducible vertices however are represented in their "channel native" frequency parametrization. After applying the parametrization changes and collecting the spin components, the final expressions are linear combinations of different spin components and frequency shifted arguments. For $F_d$ see Eq. \eqref{eq:parquet-density} and for the remaining three spin combinations we have
\begin{subequations}
\begin{align}
    F_m^{\nu \nu' \omega} &= \Lambda_m^{\nu \nu' \omega} +     \Phi_m^{\nu \nu' \omega} -\tfrac{1}{2} \Phi_d^{\nu (\nu+\omega) (\nu'-\nu)} + \tfrac{1}{2} \Phi_m^{\nu (\nu+\omega) (\nu'-\nu)}
    - \tfrac{1}{2} \Phi_s^{\nu \nu' (-\omega -\nu -\nu')} + \tfrac{1}{2} \Phi_t^{\nu \nu' (-\omega -\nu -\nu')} \,,\\
    F_s^{\nu \nu' \omega} &= \Lambda_s^{\nu \nu' \omega} + \Phi_s^{\nu \nu' \omega}+\tfrac{1}{2} \Phi_d^{\nu \nu' (-\omega-\nu-\nu')} - \tfrac{3}{2} \Phi_m^{\nu \nu' (-\omega-\nu-\nu')}
    + \tfrac{1}{2} \Phi_d^{\nu (-\nu'-\omega) (\nu' -\nu)} - \tfrac{3}{2} \Phi_m^{\nu (-\nu'-\omega) (\nu' -\nu)} \,,\\
    F_t^{\nu \nu' \omega} &= \Lambda_t^{\nu \nu' \omega} + \Phi_t^{\nu \nu' \omega} + \tfrac{1}{2} \Phi_d^{\nu \nu' (-\omega-\nu-\nu')} + \tfrac{1}{2} \Phi_m^{\nu \nu' (-\omega-\nu-\nu')}
    - \tfrac{1}{2} \Phi_d^{\nu (-\nu'-\omega) (\nu' -\nu)} - \tfrac{1}{2} \Phi_m^{\nu (-\nu'-\omega) (\nu' -\nu)} \,.
\end{align}
\label{eq:parquet-triplet}
\end{subequations}
\end{widetext}

The origin of the need for frequency shifts lies in the inherent incompatibility of the parquet equation viewpoint and the BSE viewpoint. In the BSE we choose the frequency and spin parametrizations so that we can eliminate at least one frequency and spin sum. This optimal parametrization is however different for generating \ph- and \pp-reducible diagrams. In the parquet equation we need on the other hand all vertices in the same frequency parametrization, hence the need for frequency channel transformations. 

Additionally, we also need a transformation between \ph\ and \phbar\ representations to obtain $\Phi_{\phbar}$. This transformation exploits the so called crossing symmetry relation between the \ph\ and \phbar\ frequency channels.        

\subsection{Parquet picture}

In order to have a closer look from where the frequency shifts originate, let us first extend the frequency dependence of each vertex by including a fourth fermionic Matsubara frequency $\nu_4$, i.e., a frequency related to the fourth time variable in Eq. \eqref{eq:G2}. It would multiply the time $0$ in the exponent so it is obviously redundant and given by the energy conservation $\nu_1+\nu_2+\nu_3+\nu_4=0$. Let us reintroduce the four index notation for the spin variable and use the following combined notation (as in e.g. Ref.~\onlinecite{Bickers04}):
\begin{align}
    F(1,2,3,4) = F_{\sigma_1 \sigma_2 \sigma_3 \sigma_4} (\nu_1, \nu_2, \nu_3, \nu_4).
\end{align}
Then the parquet equation \eqref{eq:parquet} is simply
\begin{align}
    F(1,2,3,4) &= \Lambda (1,2,3,4) + \Phi^{\ph} (1,2,3,4) \nonumber \\
    &\quad + \Phi^{\phbar}(1,2,3,4) + \Phi^{\pp} (1, 2, 3, 4).
\end{align}
In this notation, the crossing symmetry of the full vertex is simply 
\begin{align}
    F(1,2,3,4) =    - F(1,4,3,2) =     -F(3,2,1,4)
\end{align}
and corresponds to exchanging variables of the two creation (annihilation) operators in the expectation value in  Eq. \eqref{eq:G2} (in the language of diagrams one calls it exchanging two incoming or two outgoing lines). One can show that the reducible vertex in the \pp\ channel is also crossing-symmetric (and hence also the irreducible since $\Gamma=F-\Phi$). The crossing symmetry transformation applied to the \ph\ channel however gives only the following relation:
\begin{align}
 \Phi^{\phbar}(1,2,3,4) = -\Phi^{\ph}(1,4,3,2). 
 \label{eq:crossing-symmetry}
\end{align}

In the four frequency notation the BSEs have the following form
\begin{subequations}
\begin{align}
F(1,2,3,4) &= \Gamma^{\ph}(1,2,3,4) + \Phi^{\ph}(1,2,3,4), \\
\Phi^{\ph}(1,2,3,4)& = \Gamma^{\ph}(1,2,5,6)G(6,7)G(8,5)F(7,8,3,4), \nonumber \\
F(1,2,3,4) &= \Gamma^{\pp}(1,2,3,4)  + \Phi^{\pp}(1,2,3,4), \\
  \Phi^{\pp}(1,2,3,4)&= \tfrac{1}{2}\Gamma^{\pp}(1,5,3,6)G(6,7)G(5,8)F(7,2,4,8), \nonumber
\end{align}
\label{eq:bse1234}\end{subequations}
where the summation over repeated arguments ($5,6,7$, and $8$) is implied and we also used a two-frequency two-spin notation for the one-particle Green's function $G(1,2) = G_{\sigma_1\sigma_2}(\nu_1,\nu_2)$. This representation reveals the true difference between the \ph\ and \pp\ BSEs\textemdash{}the Green's functions connect the vertices differently, i.e., different frequency arguments are summed over. Practical evaluation of these equations requires however the introduction of two different three-frequency (two fermionic, one bosonic) parametrizations. These parametrizations are sometimes called \emph{channel native}.

\subsection{Bethe--Salpeter picture / Channel native description}

Following Ref.~\onlinecite{Thunstroem2018}, in this work we use the following frequency convention for the \ph\ and \pp\ channels
\begin{align}
    \textbf{ph: } &\nu_1 = -\nu  \qquad \qquad  &\textbf{pp: } \nu_1 &= -\nu \\
    &\nu_2 = \nu+\omega   &\nu_2 &= -(\nu'+\omega) \nonumber \\
    &\nu_3 = -(\nu'+\omega)  &\nu_3 &= \nu + \omega \nonumber \\
    &\nu_4 = \nu'  &\nu_4 &= \nu' \nonumber
\end{align}
Applying the above parametrizations to Eqs. \eqref{eq:bse1234} and additionally introducing the $d/m/s/t$ spin combinations leads to  Eqs. \eqref{eq:bse}. Now, however, we need channel transformations (frequency shifts) to evaluate the parquet equation. We need $F$ both in the \ph\ and \pp\ notations for the $d/m$ and $s/t$ channels, respectively. The channel transformations can be derived by going back and forth from three frequency to four frequency representations, e.g. 
\begin{align}
 &F_{\pp}(\nu,\nu',\omega) = F_{\pp}(-\nu_1,\nu_4,\nu_1+\nu_2) \\
    &\quad=F_{\ph} (-\nu_1,\nu_4,-\nu_2-\nu_4) = F_{\ph} (\nu,\nu',-\omega-\nu-\nu'),
    \nonumber
\end{align}    
from which we deduce
\begin{align}
    \ph  &\longrightarrow \pp\ \\
    (\nu,\nu',\omega) & \longrightarrow  (\nu, \nu',-\omega -\nu -\nu'). \nonumber 
\end{align}
To use the crossing symmetry relation \eqref{eq:crossing-symmetry}, we also need the \ph\ to \phbar\ channel transformation. The crossing transformation means exchanging either first and third or second and fourth frequency, so we can write 
\begin{align}
& F_{\phbar} (\nu,\nu',\omega) = F_{\phbar} (-\nu_1,\nu_4,\nu_1+\nu_2) = \\ 
    &\quad= F_{\ph} (-\nu_1,\nu_2,\nu_1+\nu_4)= F_{\ph} (\nu,\nu+\omega,\nu'-\nu)  \nonumber 
\end{align}
from which we deduce
\begin{align}
    \ph & \longrightarrow \phbar \\
    (\nu,\nu',\omega) & \longrightarrow  (\nu, \nu + \omega, \nu' -\nu). \nonumber 
\end{align}
All channel transformations needed in Eqs. \eqref{eq:parquet-density} and \eqref{eq:parquet-triplet} are outlined in Appendix \ref{app:channel-transformation}, together with their numerical implementation.


\section{\label{app:mpo-affine}Affine transformations}

An important subset of transformations on a QTT are coordinate transformations, in particular affine transformations. 
In this appendix, we show how to efficiently construct an MPO~\eqref{eq:TxyMPO} for such a transformation.

Rather than striving for full generality, we limit our discussion to the type of affine transformations needed in this paper: transformations between the native frequency representations for the \ph, \pp, and \phbar\ channels. In practice, we limit the range of Matsubara frequencies to a finite box. The frequencies within that box can be enumerated by positive integers. Upon transforming to another channel, some frequencies will be mapped to lie outside the frequency box of the new channel, causing missing information in the mapping. Those frequency points then have to be dropped (open boundary conditions) or periodically continued (periodic boundary conditions). With open boundary conditions, the mapping will generically become non-invertible. For the remaining frequencies, the channel transformation maps one constrained set of positive integers to another. 

We formalize the above scenario as follows.
Let $\vec x$ and $\vec y$  be vectors with $N$ components.
An affine transformation is a map $\vec x \mapsto \vec y$ that can be represented as
\begin{equation}
    \vec y = A \vec x + \vec b\,,
    \label{eq:affine}
\end{equation}
where $A$ is an invertible $N\times N$ matrix. In the following, we limit our description to the case relevant for channel transforms, where all components of \(\vec x, \vec y, \vec b, A\) are integers. We further constrain ourselves to the case where $A^{-1}$ has integer components and the components of $\vec b$ are nonnegative. Given a function $g(\vec y)$, we construct a new function $f(\vec x)$ by a coordinate transformation
\begin{equation}
    f(\vec x) := g(\vec y(\vec x)) \,.
    \label{eq:affine-general}
\end{equation}
We call this type of transformation a {\it passive} affine transformation, where for a given $\vec{x}$, we define the value of the new function $f(\vec{x})$ by picking the value of the old function $g(\vec{x})$ at the transformed point $\vec{y}$. In practice, we limit \(\vec x\), \(\vec y\) and \(\vec b\) to a finite box \(S = \{0, \ldots, 2^R - 1\}^N\).
Then, some \(\vec x\) may be transformed to a \(\vec y\) outside the box, in which case the choice of periodic or open boundary conditions becomes relevant. With periodic boundary conditions, we interpret Eq.~\eqref{eq:affine} as \(\vec y \equiv A\vec x + \vec b \pmod{2^R}\), where \(\pmod{2^R}\) is to be understood component-wise. With open boundary conditions, we set $f(\vec x) = 0$ if \(\vec y \notin S\).
The transformation \eqref{eq:affine} is not necessarily invertible on \(S\), even if it is invertible on \(\mathbb{Z}^N\).

\begin{figure}
    \centering
    \includegraphics{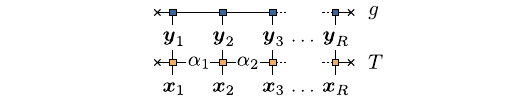}
    \caption{
        Affine transform \(T(\vec y, \vec x)\) applied to function \(g(\vec x)\) in MPO form.
    }
    \label{fig:mpo-affine}
\end{figure}

We can write Eq.~\eqref{eq:affine-general} as a tensor product
\begin{equation}
    f(\vec x) = \sum_{\vec y \in S} T(\vec x, \vec y) g(\vec y) \,,
    \label{eq:fTG}
\end{equation}
with
\begin{equation}
    T(\vec x, \vec y) :=
    \begin{cases}
        1 & \vec y = A \vec x + \vec b \,,\\
        0 & \text{else.}
    \end{cases}
    \label{eq:Tyx}
\end{equation}
In quantics representation, the tensor \(T\) can be factorized to an MPO with small bond dimension, which allows cheap transformation of functions given in QTT format through a single MPO-MPS contraction.
%
%
%
%
To construct this MPO, it is useful to start with \emph{fused} rather than interleaved indices, i.e., we only separate out the length scales of the vector, but not its components:
\begin{equation}
    \vec x = \sum_{r=1}^R 2^{R-r} \vec x_r,
\end{equation}
where $\vec x_r$ is a vector of $N$ bits corresponding to the current scale, i.e., $\vec x_r \in \{0,1\}^N$. Thus, the legs $\vec x_r$ of the corresponding MPS are of dimension $2^N$ rather than $2$. We perform similar decompositions for $\vec y$ and $\vec b$.
Consequentially, $T$ is decomposed as
\begin{equation}
    T(\vec x, \vec y) = \sum_{\alpha_1=1}^{D_1}\cdots\!\!\sum_{\alpha_{R-1}=1}^{D_{R-1}} [T_1]^{\vec x_1\vec y_1}_{1\alpha_1}
    [T_2]^{\vec x_2\vec y_2}_{\alpha_1\alpha_2} \cdots
    [T_R]^{\vec x_R\vec y_R}_{\alpha_{R-1}1},
    \label{eq:TxyMPO}
\end{equation}
where $[T_r]^{\vec x_r \vec y_r}_{\alpha_{r-1}\alpha_r}$ is the $r$-th core tensor with virtual indices $\alpha_{r-1}$ and $\alpha_{r}$ as well as local indices $\vec x_r$ and $\vec y_r$. The corresponding tensor network diagram is shown in Fig.~\ref{fig:mpo-affine}.
The indices are bound by $\alpha_r\in\{1,\ldots,D_r\}$, $\vec x_r, \vec y_r\in\{0,1\}^N$.
Once the MPO is constructed in this way, we can transform it to the interleaved representation by splitting the core tensors using a $QR$ decomposition.

For the explicit construction of the MPO, 
we first decompose Eq.~\eqref{eq:affine} for the finest scale $r=R$:
\begin{equation}
    2\vec c_{R-1} + \vec y_R = A\vec x_R + \vec b_R \,,
    \label{eq:zR}
\end{equation}
where 
$\vec c_{R-1}$ is the carry, a vector of integers not confined to 0 and 1.
Since all components of \(2\vec c_{R-1}\) are even, we find that legal values of $\vec y_R$ must satisfy:
\begin{subequations}
\begin{align}
    \vec y_R &\equiv A\vec x_R + \vec b_R   \pmod 2,
    \label{eq:yR}
\intertext{
where $(\mathrm{mod}\ 2)$ is to be understood component-wise. Consequently, the carry is obtained as
}
    \vec c_{R-1} &= \tfrac12 (A \vec x_R + \vec b_R - \vec y_R).
    \label{eq:cR}
\end{align}%
\label{eq:ycR}%
\end{subequations}%
The carry $\vec c_{R-1}$ enters the calculation for the next scale $R-1$, so we have to ``communicate'' it to the previous core tensor $T_{R-1}$ via the bond. To do so, we first observe that Eqs.~\eqref{eq:ycR} uniquely determine $\vec y_R$ and \(\vec c_R\) for each $\vec x_R$. We collect all distinct values of the carry for all possible inputs $\vec x_R$ into a tuple $(\vec c_{R-1,1}, \ldots, \vec c_{R-1,D_{R-1}})$. The core tensor is then given by
\begin{equation}
    [T_R]^{\vec x_R\vec y_R}_{\alpha 1} = \begin{cases}
        1 & 2\vec c_{R-1,\alpha} + \vec y_R = A\vec x_R + \vec b_R, \\
        0 & \mbox{else}.
    \end{cases}
\end{equation}

For all the other scales $r$, we must add the incoming carry $\vec c_r$ and must thus amend Eq.~\eqref{eq:zR} to
\begin{equation}
    2\vec c_{r-1} + \vec y_r = A\vec x_r + \vec b_r + \vec c_r \,,
    \label{eq:zr}
\end{equation}
and Eqs.~\eqref{eq:ycR} to:
\begin{subequations}
\begin{align}
    \vec y_r &\equiv A\vec x_r + \vec b_r + \vec c_r  \pmod 2 \,,
    \\
    \vec c_{r-1} &= \tfrac12 (A \vec x_r + \vec b_r + \vec c_r - \vec y_r).
\end{align}
\label{eq:ycr}
\end{subequations}
We again collect all distinct outgoing carry values for all possible $\vec x_r$ and $\vec c_r$ into $(\vec c_{r-1,\alpha})_{\alpha=1,\ldots,D_{r-1}}$, and obtain the core tensor
\begin{equation}
    [T_r]^{\vec x_r\vec y_r}_{\alpha\alpha'} = \begin{cases}
        1 & 2\vec c_{r-1,\alpha} + \vec y_r = A\vec x_r + \vec b_r + \vec c_{r,\alpha'}, \\
        0 & \mbox{else}.
    \end{cases}
    \label{eq:Tr}
\end{equation}

We iterate this procedure from \(r = R\) to \(1\), constructing all MPO core tensors in a single backward sweep.
Having reached the first tensor, \(r = 1\), we implement open boundary conditions by demanding that \(\vec c_0 = \vec 0\) in Eq.~\eqref{eq:Tr}, such that
\begin{equation}
    [T_1]^{\vec x_1\vec y_1}_{1\alpha'} = \begin{cases}
        1 & \vec y_1 = A\vec x_1 + \vec b_1 + \vec c_{1,\alpha'}, \\
        0 & \mbox{else}.
    \end{cases}
    \label{eq:T1open}
\end{equation}
Periodic boundary conditions are implemented by modifying Eq.~\eqref{eq:Tr} such that the leftmost carry \(\vec c_0\) is discarded, such that
\begin{equation}
    [T_1]^{\vec x_1\vec y_1}_{1\alpha'} = \begin{cases}
        1 & \vec y_1 \equiv A\vec x_1 + \vec b_1 + \vec c_{1,\alpha'} \pmod 2, \\
        0 & \mbox{else}.
    \end{cases}
    \label{eq:T1periodic}
\end{equation}

The bond dimension \(\Dmax = \max_r D_r\) of the tensors constructed in this way is likely optimal.
This algorithm has $\mathcal O(R \Dmax^2 2^{2N})$ runtime, which is optimal in the sense that at least this amount of runtime and memory is necessary to construct the tensors \(T_r\).
The algorithm can be generalized to cases where \(A\) and \(\vec b\) have entries in \(\mathbb{Q}\).



\begin{table}
    \centering

\setlength{\tabcolsep}{0.6em}
    \begin{tabular}{r|ccc|ccc}
\toprule
$r$ & $\alpha_{r-1}$ & $\vec c_{r-1}$ & $\vec y_r$ & $\vec x_r$ & $\alpha_r$ & $\vec c_r$ \\
\midrule
$R$   & $1$ & $(0, \ms0)$ & $(0, 0)$ & $(0, 0)$ & $1$ & $(0, \ms0)$ \\
      & $1$ & $(0, \ms0)$ & $(1, 1)$ & $(1, 0)$ & $1$ & $(0, \ms0)$ \\
      & $2$ & $(0,   -1)$ & $(0, 1)$ & $(0, 1)$ & $1$ & $(0, \ms0)$ \\
      & $1$ & $(0, \ms0)$ & $(1, 0)$ & $(1, 1)$ & $1$ & $(0, \ms0)$ \\
\midrule
$R-1$ & $1$ & $(0, \ms0)$ & $(0, 0)$ & $(0, 0)$ & $1$ & $(0, \ms0)$ \\
      & $1$ & $(0, \ms0)$ & $(1, 1)$ & $(1, 0)$ & $1$ & $(0, \ms0)$ \\
      & $2$ & $(0,   -1)$ & $(0, 1)$ & $(0, 1)$ & $1$ & $(0, \ms0)$ \\
      & $1$ & $(0, \ms0)$ & $(1, 0)$ & $(1, 1)$ & $1$ & $(0, \ms0)$ \\[1ex]
      & $2$ & $(0,   -1)$ & $(0, 1)$ & $(0, 0)$ & $2$ & $(0, -1)$ \\
      & $1$ & $(0, \ms0)$ & $(1, 0)$ & $(1, 0)$ & $2$ & $(0, -1)$ \\
      & $2$ & $(0,   -1)$ & $(0, 0)$ & $(0, 1)$ & $2$ & $(0, -1)$ \\
      & $2$ & $(0,   -1)$ & $(1, 1)$ & $(1, 1)$ & $2$ & $(0, -1)$ \\
\midrule
$\vdots$ & $\vdots$ & $\vdots$ & $\vdots$ & $\vdots$ & $\vdots$ & $\vdots$ \\
\midrule
$1$   & $1$ & $(0, \ms0)$ & $(0, 0)$ & $(0, 0)$ & $1$ & $(0, \ms0)$ \\
      & $1$ & $(0, \ms0)$ & $(1, 1)$ & $(1, 0)$ & $1$ & $(0, \ms0)$ \\
      & $1$ & $(0, \ms0)$ & $(1, 0)$ & $(1, 1)$ & $1$ & $(0, \ms0)$ \\[1ex]
      & $1$ & $(0, \ms0)$ & $(1, 0)$ & $(1, 0)$ & $2$ & $(0, -1)$ \\
\bottomrule
    \end{tabular}
\setlength{\tabcolsep}{2.0pt}
    \caption{%
Non-zero elements of $[T_r]^{\vec x_r \vec y_r}_{\alpha_{r-1}\alpha_r}$~\eqref{eq:TxyMPO} constructed from Eqs.~\eqref{eq:zR}--\eqref{eq:T1open} for the affine transform \eqref{eq:affine-example} with open boundary conditions.
    }
    \label{tab:affine}
\end{table}

Let us walk through the algorithm for the example transformation with $N=2$ and open boundary conditions:
\begin{equation}
    \vec y =
    \begin{pmatrix} 1 & \ms0 \\ 1 & -1 \end{pmatrix} \vec x +
    \begin{pmatrix} 0 \\ 0 \end{pmatrix}
    .\label{eq:affine-example}
\end{equation}
The non-zero elements of the corresponding core tensors~\eqref{eq:TxyMPO} are listed in Table~\ref{tab:affine}. For the case $r=R$, we simply apply the transformation \eqref{eq:yR} to each bit combination $\vec x_R \in \{(0,0), (1,0), (0,1), (1,1)\}$. In the case $\vec x_R = (0,1)$, Eq.~(\ref{eq:cR}) yields a carry of $\vec c_{R-1} = (0,-1)$, to which we assign the bond index $\alpha_{R-1}=2$, otherwise it is $(0,0)$, to which we assign the index $\alpha_{R-1}=1$. The dimension of the corresponding bond is thus $D_{R-1} = 2$ and the core tensor has the four nonzero elemnts listed in rows 1--4 of Table~\ref{tab:affine}.

For $r=R-1$, Eq.~\eqref{eq:ycr} directs us to add the incoming carry $\vec c_r$. Hence, we double the number of non-zero entries, as we have to repeat the calculation for each of the two outgoing carries of $T_R$.  We observe that the set of incoming and outgoing carries is identical and assign the same bond indices to them. The corresponding nonzero elements of $T_{R-1}$ are then listed in rows 5--12 of Table~\ref{tab:affine}. Since the values of the outgoing carries form the same set as those of the incoming carries, all other core tensors $T_{r'}$ with $1 < r' < R$ are equal to $T_{R-1}$.
For $r=1$, we impose open boundary conditions, thereby restricting the outgoing carry of $T_1$ to zero, as shown in Eq.~\eqref{eq:T1open}. This cuts half of the elements and yields the entries listed in rows 13--16 of Table~\ref{tab:affine}.

\section{\label{app:channel-transformation}Channel transformations}
We describe how to implement channel transformations using affine transformations in QTT, which is defined in Eq.~\eqref{eq:affine-general}.

\subsection{\ph\ to \pp\ transformation}
We first describe a \ph\ $\rightarrow$ \pp\ channel transformation via the \phbar\ channel:
\begin{alignat}{3}
    & \ph && \longrightarrow \phbar && \longrightarrow \pp \\
    & (\nu,\nu',\omega) && \longrightarrow (\nu,\nu+ \omega, \nu'-\nu) && \longrightarrow (\nu, \nu',-\omega -\nu -\nu') \,,\nonumber
\end{alignat}
where $\nu^{(')} = (2n^{(')}+1)\pi /\beta$ and $\omega = 2m \pi /\beta$. The picture corresponds to the following, e.g. for the full vertex:
\begin{subequations}
\begin{align}
    F_{\pp}(\nu,\nu',\omega) &= F_{\ph} (\nu, \nu',-\omega -\nu -\nu'), \\
    F_{\phbar} (\nu,\nu',\omega) &= F_{\ph} (\nu,\nu+ \omega, \nu'-\nu), \\
    F_{\pp}(\nu,\nu',\omega) &= F_{\phbar} (\nu,-\nu'-\omega, \nu' -\nu).
\end{align}
\end{subequations}
In the following, we will denote the ``old'' variables in every transformation step with a tilde. For the \ph\ to \phbar\  transformation 
\begin{align}
    F_{\phbar} (\nu,\nu',\omega) &= F_{\ph} (\nu,\nu+ \omega, \nu'-\nu) = F_{\ph} (\tilde{\nu}, \tilde{\nu}', \tilde{\omega})
\end{align}
we need the transformation matrix (expressing the old $(\tilde{\nu}, \tilde{\nu}', \tilde{\omega})$ by the new variables $(\nu, \nu',\omega)$)
\begin{align}
    \begin{pmatrix} \tilde{\nu} \\ \tilde{\nu}' \\ \tilde{\omega} \end{pmatrix} = 
    \begin{pmatrix}
        1 & 0 & 0 \\
        1 & 0 & 1 \\
        -1 & 1 & 0
    \end{pmatrix} 
    \begin{pmatrix} \nu \\ \nu' \\ \omega  \end{pmatrix}
    \,,
\end{align}
with \((\tilde{\nu}, \tilde{\nu}', \tilde{\omega}) = (\nu,\nu+ \omega, \nu'-\nu)\) and $\tilde \nu^{(')} = (2 \tilde n^{(')}+1)\pi /\beta$ and $\tilde \omega = 2 \tilde m \pi /\beta$.
We also need to shift the indices such that, for example for $n = 0$ and $m = 0$ ($\nu + \omega = \frac{\pi}{\beta} = \tilde{\nu}'$) we are at $\tilde{n}'=0$ again.
\begin{gather}
    0 \leq a,b,c, \tilde a, \tilde b, \tilde c \leq N-1 \nonumber \\
    n = a - \tfrac{N}{2} \,, \quad
    n'= b - \tfrac{N}{2} \,, \quad
    m = c - \tfrac{N}{2} \,, \\
    \tilde n = \tilde a - \tfrac{N}{2} \,, \quad
    \tilde n'= \tilde b - \tfrac{N}{2} \,, \quad
    \tilde m = \tilde c - \tfrac{N}{2} \,,
\end{gather}
with $N=2^R$.
This leads to
\begin{alignat}{3}
    \tilde{a} &= \phantom{-} a, &&&& \textrm{no shift} \nonumber  \\
    \tilde{b} &= \phantom{-} a+b-\tfrac{N}{2} &&= \phantom{-} n+m+\tfrac{N}{2}, ~&& \textrm{shift by } \tfrac{N}{2} \nonumber  \\
    \tilde{c} &= -a+b+\tfrac{N}{2} &&= -n+n'+\tfrac{N}{2}, ~&&\textrm{shift by } \tfrac{N}{2}.
\end{alignat}
Hence, we get the shift vector $\vec{b} = (0,\frac{N}{2},\frac{N}{2})^\mathrm{T}$. For example, at $n = n'$ we need $\omega = 0$ and, thus, $\tilde{c} = \frac{N}{2}$, which is ensured by the shift. This first transformation can be represented by an MPO with $\Dmax = 9$.

For the \phbar\ $\rightarrow$ \pp\ transformation,
\begin{align}
    F_{\pp}(\nu,\nu',\omega) &= F_{\phbar} (\nu,-\nu'-\omega, \nu' -\nu) = F_{\phbar} (\tilde{\nu}, \tilde{\nu}', \tilde{\omega}),
\end{align}
we need the transformation matrix 
\begin{align}
    \begin{pmatrix} \tilde{\nu} \\ \tilde{\nu}' \\ \tilde{\omega} \end{pmatrix} = 
    \begin{pmatrix}
        1 & 0 & 0 \\
        0 & -1 & -1 \\
        -1 & 1 & 0
    \end{pmatrix}
    \begin{pmatrix} \nu \\ \nu' \\ \omega  \end{pmatrix},
\end{align}
with \((\tilde{\nu}, \tilde{\nu}', \tilde{\omega}) = (\nu,-\nu'- \omega, \nu'-\nu)\).
Using the same procedure as above, we get the shift vector \mbox{$\vec{b} = (0,\frac{N}{2}-1,\frac{N}{2})^\mathrm{T}$}. This affine transformation can be represented by an MPO with $\Dmax = 15$.
%

\subsection{\pp\ to \ph\ transformation}
Basically, we use the same procedure as above since the \ph\ to \pp\ transformation is its own inverse. 
\begin{alignat}{3}
    & \pp && \longrightarrow \overline{\pp} && \longrightarrow \ph\\
    & (\nu,\nu',\omega) && \longrightarrow (\nu,\nu+ \omega, \nu'-\nu) && \longrightarrow (\nu, \nu',-\omega -\nu -\nu'), \nonumber 
\end{alignat}
The picture corresponds to the following, e.g. for the full vertex
\begin{subequations}
\begin{align}
    F_{\ph}(\nu,\nu',\omega) &= F_{\pp} (\nu, \nu',-\omega -\nu -\nu'), \\
    F_{\overline{\pp}} (\nu,\nu',\omega) &= F_{\pp} (\nu,\nu+ \omega, \nu'-\nu), \\
    F_{\ph}(\nu,\nu',\omega) &= F_{\overline{\pp}} (\nu,-\nu'-\omega, \nu' -\nu).
\end{align}
\end{subequations}
The transformation matrices and shift factors and, hence, the MPO representations are identical to the \ph\ to \pp\ transformation.

\section{\label{app:qtt-parquet}Parquet equation in QTT format}
\begin{figure}
    \centering 
    \includegraphics[width=\linewidth]{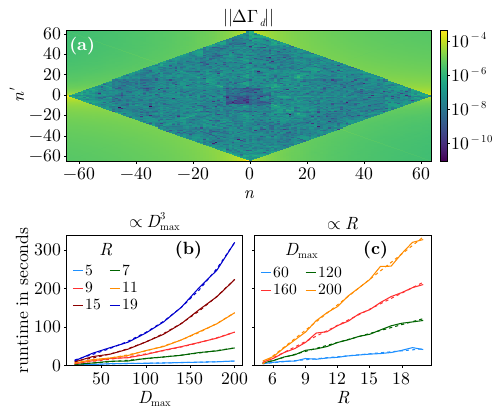}
    \caption{(a) Absolute normalized error $||\Delta \Gamma_{d}|| := |\Gamma_{d, \mathrm{parquet}} - \Gamma_{d, \mathrm{exact}}|/||\Gamma_{d, \mathrm{exact}}||_{\infty}$ of the parquet equation with QTTs for $\Gamma_d$ compared to the exact values at $\omega=0$ in the fermionic Matsubara frequency plane for $\Dmax=200$, $R=7$, $\beta = U=1$, $\epsilon = 10^{-8}$. Runtime of the parquet equation for $\Gamma_d$ for various maximum bond dimensions and grid size parameters $R$ with $\beta=U=1$. The dashed lines indicate (b) the cubic runtime increase with $\Dmax$ and (c) linear increase by increasing R, which corresponds to an exponential increase in the number of grid points.}
    \label{fig:parquet-error}
\end{figure}

In the parquet equation \eqref{eq:parquet-density} and \eqref{eq:parquet-triplet} the triangle-shaped frequency box errors from frequency transformations add up to a diamond shaped error with larger errors in the corners. This can be observed in Fig.~\ref{fig:parquet-error}, where a plot of the absolute normalized error of the irreducible vertex in the density channel $\Gamma_d$ computed via Eq.~\eqref{eq:parquet-density} and \eqref{eq:phi} with QTCI compared to the exact $\Gamma_d$ is shown at $\omega=0$. 

The cubic dependence on $\Dmax$ is shown in Fig.~\ref{fig:parquet-error}(b), which corresponds to the cubic scaling of the channel transformations inside the parquet equations and, thus, constitutes the bottleneck of the parquet equation.
Linear scaling of the runtime of the parquet equation with QTTs in $R$ is shown in Fig.~\ref{fig:parquet-error}(c).
Again, we want to emphasize that exponentially increasing the number of grid points comes only at linearly increasing computational cost in the parquet equation.

\section{\label{app:qtt-sde} SDE in QTT format}
\begin{figure}
    \centering 
    \includegraphics[width=\linewidth]{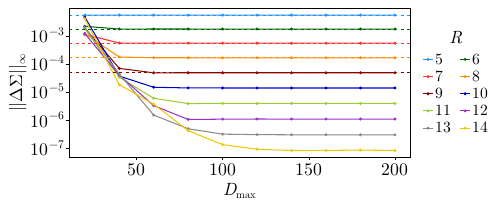}
    \caption{Plot of the maximum absolute normalized error of the SDE $||\Delta \Sigma||_{\infty} := ||\Sigma_{d,\mathrm{SDE}} - \Sigma_{d,\mathrm{exact}}||_{\infty}/||\Sigma_{d,\mathrm{exact}}||_{\infty}$ with QTTs
    for the self-energy $\Sigma$ compared to the exact values for various grid sizes depending on the maximum bond dimension with $U=\beta=1$. The dashed lines indicate the results from the dense grid calculation without QTTs.} 
    \label{fig:sde-error}
\end{figure}

In Fig.~\ref{fig:sde-error}, we show the maximum absolute normalized error of the self-energy $\Sigma$ obtained by using QTTs in the SDE. The dashed lines represent the errors of the dense grid calculations, which are due to the finite size of the grid. A qualitatively similar behavior to the BSE can be observed, with the difference that already quite low bond dimensions are sufficient for obtaining very low errors. This is caused by the frequency dependence of the functions in the SDE, where only the full vertex depends on three Matsubara frequencies. 
At this point, we should repeat emphasizing the fact that exponentially increasing the number of grid points by increasing the grid parameter $R$ exponentially reduces 
the error (exponential convergence to the true solution), but only comes with linearly increasing computational cost. In the case of $R=14$ the computation with a maximum bond dimension of 100 took only around 100 seconds using QTTs on a single 512 GB node on a cluster, while without the use of QTTs the calculation would include computations with the numerical data of the full vertex, which is the size of $8 \times 2^{3\times 14} \simeq 3.5\times 10^{13}$ bytes. This would only be possible by engaging a larger number of nodes on a cluster, which emphasizes the strength of the QTCI approach.

\section{\label{app:technical-limitations}Technical limitations}
Theoretically, in the iterative parquet calculations with QTCI, it should easily be possible to run calculations for much larger grids, e.g. $R = 20$ ($2^{3\times 20}$ grid points), on a single 512 GB node on a cluster, without running into any memory or computational time bottlenecks, because the computational costs only depend linearly on $R$. This is still true, but there is another limiting technical difficulty at the moment.

In the calculations, a specified maximum bond dimension is set not only in the initial TCI, but also in every QTT operation, in order to avoid a blowing up of the bond dimensions since this is the computational bottleneck. As was shown in Sec.~\ref{sec:tci-compression}, for the initial TCI already bond dimensions slightly above 100 are sufficient to reach maximum normalized errors of $10^{-6}$ of the QTTs with respect to the exact data.
However, a problem emerges in the QTT operations,
which are at the moment SVD based and make use of the \texttt{truncate} function in \texttt{ITensors.jl} to compress the resulting QTTs back to a certain maximum bond dimension. Furthermore, the fit algorithm used for MPO-MPO contractions relies on the SVD truncation internally~\cite{Verstraete2004}.
The SVD truncation minimizes the difference between an original MPS and an approximated one in terms of the Frobenius norm.
Because the Frobenius norm of the vertex functions grows exponentially with $R$ due to a constant term, the SVD truncation is expected to fail at large $R$; the Frobenius norm reaches $c (2^R)^3 \approx c\times 3\times 10^{13}$ at $R=15$ ($c$ is the constant term).

\begin{figure}[!h]
    \centering 
    \includegraphics[width=\linewidth]{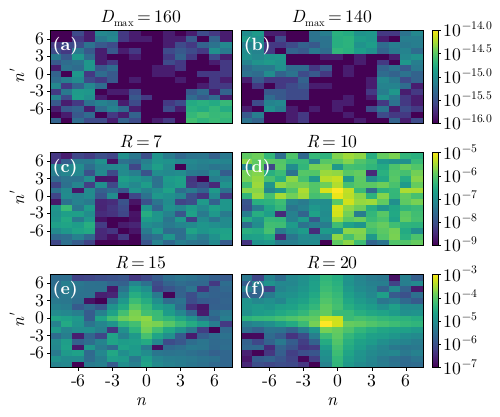}
    \caption{Absolute normalized error of the reconstructed $F_d$ compared to the exact data for the innermost $16 \times 16$ fermionic Matsubara frequency grid at $\omega = 0, \beta=U=1$. (a) and (b) show the error after using TCI with a set tolerance of $10^{-10}$ and the maximum bond dimensions set to 160 and 140 respectively. (c)--(f) show the error after applying the SVD based truncation to the QTT with maximum bond dimension 160 (a) with a set maximum bond dimension of 140. The truncation error increases with larger values of the grid parameter $R$.
    }
    \label{fig:tci-error-truncation}
\end{figure}

Here, we have observed the truncation error to become more significant the larger the value of the grid parameter $R$ is and even leads to wrong results around $R=15$. 
In Fig.~\ref{fig:tci-error-truncation}, we show this behavior in case of the full vertex in the density channel $F_d$.
In Figs.~\ref{fig:tci-error-truncation}(a-b) the absolute normalized error is shown after applying TCI to evaluate a QTT for $F_d$ for different maximum bond dimensions. It can be seen that the error in this center ($16 \times 16$) fermionic Matsubara frequency box is of $\order(10^{-15})$.
Figures~\ref{fig:tci-error-truncation}(c-f) show the error after applying the SVD based truncation to the QTT with maximum bond dimension 160 down to a maximum bond dimension of 140 for various grid sizes determined by $R$. Although using TCI with a maximum bond dimension of 140 [Fig.~\ref{fig:tci-error-truncation}(b)] the QTT was able to reconstruct the exact data with an absolute normalized error of $\order(10^{-15})$, applying the SVD based truncation significantly worsens the results, leading to normalized errors between \(10^{-9}\) and \(10^{-5}\).
Moreover, it can be seen that the error increases significantly with increasing $R$. This is why in the case of the iterative parquet solutions for the Hubbard atom, only results up to $R=11$ are shown, since the resulting maximum normalized error does not improve anymore for larger grids due to the truncation errors. This can also already seen in Fig.~\ref{fig:iterative-parquet-sde-error} for $R=11$, where the maximum normalized error at a maximum bond dimension of 200 is only slightly lower than in the case of $R=10$.
However, this should only be a problem of the current implementation and first numerical tests indicate that it is possible to overcome this limitation in the future e.g. by using CI based truncation~\cite{NunezFernandez2024}. This is because the CI-based truncation relies on the maximum norm and thus does not suffer from the divergence of the Frobenius norm. 

An alternative way to deal with the infinite Frobenius norm is, as mentioned at the end of Sec.~\ref{sec:results}, to change into formalism with vertices with removed asymtotics and thus finite Frobenius norm. The recent reformulation of parquet equations into single- and multi-boson exchange vertices provides such a  solution~\cite{Krien2020b,Krien2022}. Earlier approaches to parquet equations have used the so-called kernel asymptotics~\cite{victory2019,Wentzell2020}.

\section{\label{app:extensions}Model extensions}

In future work, our goal will be to apply the QTT representation to two-particle calculations with orbital and momentum degrees of freedom. Suppose that these are labeled by an additional (composite) index, say $i = 1, \dots, N$,
then the vertex carries four such indices, $F_{ijkl}$, and has $N^4$ components.
The memory costs for such computations depend on how QTTs are used to parametrize the vertex.

For example, a naive approach would be to use a separate QTT to parameterize the frequency dependence of 
$F_{ijkl} (\nu, \nu', \omega)$
for each index combination $(i,j,k,l)$.
This would require $N^4$ different QTTs and $N^4$ BSEs connecting them all, etc. We estimate that with this approach, 
computations for $N=6$, $R=10$ and $\Dmax = 200$ should be feasible on a single 512 GB node.

However, such a naive approach would not exploit low-rank structures that may  arise if different vertex components have similar frequency dependencies. In such a case, it could be more efficient to use a single QTT to parametrize the dependence of the vertex on its frequencies \textit{and} all $i$-indices. To pursue such a strategy and optimize its efficiency,  further  methodological developments will be required 
to address some open questions: 
What is the best grouping and order of quantics indices for a combined frequency and momentum/orbital representation? How can MPO-MPO contractions (the current bottleneck) be performed more efficiently? 
What are the best strategies for parallelizing the computations? These issues are currently being explored in ongoing work.

\bibliography{Bibliography_QTTparquet}
\end{document}